\documentclass[twocolumn, times]{aastex63}

\usepackage{graphicx}
\usepackage{amsfonts,amsmath,amssymb}
\usepackage{xspace}
\usepackage{bm}

\newcommand{\brunt}{Brunt-V\"ais\"al\"a\xspace}
\newcommand{\Msun}{\ensuremath{\mathrm{M}_\odot}\xspace}
\newcommand{\vtrb}{\ensuremath{v_\mathrm{turb}}\xspace}
\newcommand{\Lamm}{\ensuremath{\Lambda_\mathrm{mix}}\xspace}
\newcommand{\alphL}{\ensuremath{\alpha_\Lambda}\xspace}
\newcommand{\Ye}{\ensuremath{Y_e}\xspace}

\received{\today}
\submitjournal{ApJ}

\shorttitle{Turbulence-aided Supernovae}
\shortauthors{Couch, Warren, \& O'Connor}

\begin{document}

\title{\large Simulating Turbulence-aided Neutrino-driven Core-collapse Supernova Explosions in One Dimension}

\author[0000-0002-5080-5996]{Sean M.~Couch}
\affiliation{Department of Physics and Astronomy, Michigan State University, East Lansing, MI 48824, USA}
\affiliation{Department of Computational Mathematics, Science, and Engineering, Michigan State University, East Lansing, MI 48824, USA}
\affiliation{National Superconducting Cyclotron Laboratory, Michigan State University, East Lansing, MI 48824, USA}
\affiliation{Joint Institute for Nuclear Astrophysics-Center for the Evolution of the Elements, Michigan State University, East Lansing, MI 48824, USA}

\author[0000-0001-9440-6017]{MacKenzie L.~Warren}
\altaffiliation{NSF Astronomy \& Astrophysics Postdoctoral Fellow}
\affiliation{Department of Physics and Astronomy, Michigan State University, East Lansing, MI 48824, USA}
\affiliation{Joint Institute for Nuclear Astrophysics-Center for the Evolution of the Elements, Michigan State University, East Lansing, MI 48824, USA}

\author[0000-0002-8228-796X]{Evan P. O'Connor}
\affiliation{Department of Astronomy and The Oskar Klein Centre, Stockholm
  University, AlbaNova, SE-106 91 Stockholm, Sweden}

\begin{abstract}
  The core-collapse supernova (CCSN) mechanism is fundamentally three-dimensional with instabilities, convection, and turbulence playing crucial roles in aiding neutrino-driven explosions.
  Simulations of CCNSe including accurate treatments of neutrino transport and sufficient resolution to capture key instabilities remain amongst the most expensive numerical simulations in astrophysics, prohibiting large parameter studies in 2D and 3D.
  Studies spanning a large swath of the incredibly varied initial conditions of CCSNe are possible in 1D, though such simulations must be artificially driven to explode.
  We present a new method for including the most important effects of convection and turbulence in 1D simulations of neutrino-driven CCSNe, called Supernova Turbulence In Reduced-dimensionality, or {\it STIR}.
  Our new approach includes crucial terms resulting from the turbulent and convective motions of the flow.
  We estimate the strength of convection and turbulence using a modified mixing length theory (MLT) approach introducing a few free parameters to the model which are fit to the results of 3D simulations.
  For sufficiently large values of the mixing length parameter, turbulence-aided neutrino-driven explosions are obtained. 
  We compare the results of STIR to high-fidelity 3D simulations and perform a parameter study of CCSN explosion using 200 solar-metallicity progenitor models from 9 to 120 \Msun. 
  We find that STIR is a better predictor of which models will explode in multidimensional simulations than other methods of driving explosions in 1D.
  We also present a preliminary investigation of predicted observable characteristics of the CCSN population from STIR, such as the distributions of explosion energies and remnant masses.
\end{abstract}


\section{Introduction}

Core-collapse supernovae (CCSNe) are the explosive deaths of stars more massive than about 8 \Msun.
The connection between massive stars and CCSNe is now well-established. 
Scores of direct progenitor identifications have been made from archival imaging \citep{smartt:2009, smartt:2015, van-dyk:2012a, van-dyk:2012b, van-dyk:2013} dating back to the famous case of SN 1987A \citep{sonneborn:1987}.
While there is increasing observational certainty that CCSNe arise from massive stars, our theoretical understanding of the mechanism that drives these explosions is still incomplete.

Massive stars reach temperatures and densities in their cores sufficient to synthesize iron.
These iron cores are inert and, thus, the end point of stellar nuclear fusion. 
Continued nuclear ``burning'' builds up the iron cores to the effective Chandrasekhar mass \citep{baron:1990} and gravitational instability and collapse ensues.
The collapse accelerates until nuclear density is exceeded at which point the strong nuclear force becomes, quite suddenly, repulsive.
The collapse is halted in a matter of milliseconds, launching a strong shock wave into the still-collapsing mantle of the core in a process known as core ``bounce.''
Electron captures on iron-group nuclei during the collapse leave the now-quasi-hydrostatic inner core composed mostly of neutrons, a proto-neutron star (PNS).
Neutrinos are also ubiquitous after core bounce, being produced by both electron and positron captures as well as thermal processes like electron-positron annihilation.  
They ultimately carry away the vast majority of the gravitational binding energy released in the collapse, well over $10^{53}$\,erg.

The shock created by core bounce moves out quickly at first but loses energy to dissociation of iron nuclei and precipitous neutrino cooling of the post-shock medium. 
The shock ultimately stalls, typically around 150 km in radius, above the nascent PNS transitioning into an accretion shock.
Understanding the mechanism that revives the outward motion of the shock and supplies the energy necessary to unbind the envelope of the progenitor star has been a long-standing problem in theoretical astrophysics.
For reviews of the quest to understand the CCSN mechanism, see \citet{bethe:1990, janka:2007, janka:2012a, janka:2016, janka:2012, burrows:2013, muller:2016b, couch:2017}. 

The modern paradigm for the CCSN explosion mechanism is the neutrino-heating mechanism \citep{bethe:1985, bruenn:1985}, first proposed by \citet{colgate:1966} and \citet{arnett:1966}.
The key idea is that neutrinos liberated during the post-bounce accretion phase can heat the region behind the shock sufficiently to initiated shock re-expansion and explosion. 
Neutrino heating in the so-called ``gain'' layer behind the shock, where neutrino heating exceeds neutrino cooling, is very inefficient and neutrino-driven explosions have been notoriously hard to come by, particularly in spherical symmetry \citep[cf.][]{arnett:1966, bruenn:1985, liebendorfer:2001}.
Throughout its history the neutrino mechanism has been beset by significant uncertainties in key physics, such as the equation of state (EOS) of nuclear material and neutrino-matter interactions.
Our physical understanding of both the nuclear EOS \citep{lattimer:2012, steiner:2013, hebeler:2013} and key neutrino interactions \citep{burrows:2006a, horowitz:2002, roberts:2012b, bollig:2017, burrows:2018}, however, has advanced significantly, clearing the way for a modern, predictive theory of the neutrino mechanism.

Tremendous progress has been made in our understanding of the CCSN mechanism in recent years, spurred largely by the emerging capability for high-fidelity simulations in 3D \citep[cf.][]{janka:2016, muller:2017, andresen:2017, summa:2018, lentz:2015, oconnor:2018b, vartanyan:2019}.
Such simulations are extremely challenging, requiring high-resolution (magneto-)hydrodynamics, general relativistic gravity, a complex microphysical EOS, and accurate neutrino transport.
This latter requirement is typically the stiffest challenge, and greatest expense, in modern CCSN mechanism simulations. 
3D CCSN simulations using non-parametric approaches to neutrino transport can cost millions of node-hours on modern supercomputers {\it per simulation}.
This limits our ability to carry out large parameter studies of the CCSN mechanism in 3D.

CCSNe arise from an enormous variety of initial conditions.
The parameter space for CCSN progenitors includes dimensions of zero-age main sequence (ZAMS) mass, metallicity, rotation rate, and even binary system parameters such as companion mass and separation. 
Each of these variables can have a significant impact on the progenitor structure and, hence, CCSN simulations and the resulting predictions for key observables.  
Additionally, uncertainties in key microphysical inputs, such as the nuclear EOS and neutrino-matter interactions, may also lead to significant impacts on the results of CCSN simulations.
This parameter space is certainly too large to explore in 3D at present, and so population studies of CCSNe have only really been carried out in 1D \citep[e.g.,][]{ugliano:2012, pejcha:2015, sukhbold:2016, muller:2016, fischer:2018, ebinger:2019}. 
The drawback, however, is that we know that the CCSN mechanism is fundamentally multidimensional \citep[e.g.,][]{marek:2009a, muller:2012, bruenn:2013,bruenn:2016, summa:2016, burrows:2018, oconnor:2018, oconnor:2018b} and the notorious difficulty of obtaining explosions in 1D necessitates some artificial means of driving explosions for such studies.

Some of the earliest, and still popular, means for exploding massive stars in 1D were ``pistons'' \citep[e.g.,][]{woosley:1995} and ``thermal bombs'' \citep[e.g.,][]{nomoto:2006}. 
In the former, an inner Lagrangian boundary is contracted, simulating the collapse of the iron core, then quickly expanded, launching a strong shock into the collapsing star that drives explosions. 
The motion of this inner boundary, which is set by hand, determines the character of the explosions and resulting observables, including the nucleosynthesis. 
In the case of thermal bombs, the explosions are driven by artificially heating the matter in the core of the star over a brief period of time. 
In both models, the ``mass cut'', or equivalently the mass of the compact remnant neglecting fallback, is set by hand and can impact the results significantly. 
Critically, both of these popular approaches to driving 1D explosions neglects physics we know to be crucial to the CCSN mechanism, specifically neutrino transport and a microphysical EOS. 

Recently, a few groups have developed models for driving 1D CCSN explosions that do include neutrino physics and realistic equations of state. 
In \citet{ugliano:2012}, the authors present a 1D explosion model that utilizes gray neutrino transport \citep[cf.,][]{scheck:2006} and a contracting inner boundary that mimics the contracting PNS. 
The rapidity of the PNS contraction, and hence rate of gravitational binding energy liberated, sets the neutrino luminosity and, therefore, heating in the gain layer behind the stalled shock. 
For sufficiently rapid contraction of this inner boundary, 1D neutrino-driven explosions are obtained. 
Studies using this approach \citep{sukhbold:2016, ertl:2016} show encouraging agreement with certain features of the observed CCSN population.
As with pistons and thermal bombs, the character of these explosions is sensitive to the nature of the imposed parameters of the model. 
One concern with this approach is that the electron-type neutrino luminosities may be enhanced relative to, or at least substantially different from, those in multidimensional simulations.
This could have a significant impact on the resulting nucleosynthesis since the electron-type neutrino luminosities set the electron fraction in the ejecta.

An alternative means of simulating neutrino-driven explosions in 1D is presented by \citet{perego:2015}.
These authors present the ``PUSH'' model for explosions that relies on an artificial additional neutrino heating source that depends on the luminosities of the heavy-lepton neutrinos, which realistically contribute negligibly to the neutrino heating.
This model includes the full core of the PNS and so is better able to address questions regarding the impact of the microphysical EOS and, since it avoids directly altering the luminosities of the electron-type neutrinos, is better suited to studies of nucleosynthesis from 1D CCNSe \citep{ebinger:2019, curtis:2019}.
Explosions achieved with PUSH are also, however, sensitive to the free parameters of the artificial heating model used and the predicted observables will vary according to the character of the artificially augmented neutrino heating.
PUSH uses the isotropic diffusion source approximation \citep[IDSA][]{liebendorfer:2009} for electron-type neutrino transport and a parameterized leakage scheme \citep{oconnor:2010} for the heavy-lepton neutrinos.
Recent controlled code-to-code comparisons in 1D \citep{oconnor:2018a}  and 2D \citep{pan:2019} have shown that IDSA gives slightly different answers as compared to higher-fidelity transport methods.
For both PUSH and the models of \citet{ugliano:2012, sukhbold:2016}, the free parameters of the explosion model are chosen on the basis of fitting certain observational parameters of real CCSNe, such as SN 1987A or the Crab.

Approximate methods for driving 1D explosions have tremendous value in allowing the exploration of simulated CCSN {\it populations}, as well as the specific details of individual models. 
Still, it is not clear how faithfully these 1D models reproduce the results of high-fidelity multidimensional simulations, even when population statistics such as the mean explosion energy and remnant masses compare well to that of the observed population.
In particular, there is some tension between the predicted explodability of progenitors from the 1D models of \citet{sukhbold:2016} and the 2D high-fidelity simulations of \citet{oconnor:2018}. 
In \citet*{oconnor:2018}, we present 2D simulations of progenitor stars with ZAMS masses of 12-, 15-, 20-, 21-, 22-, 23-, 24-, and 25-\Msun. 
All of these progenitors explode in those simulations except for the 12- and 21-\Msun stars. 
This is precisely the opposite behavior as found in \citet{sukhbold:2016} for these same progenitor masses.\footnote{We note that in \citet{oconnor:2018} the progenitor models of \citet{woosley:2007} were used where as in \citet{sukhbold:2016} the model set of \citet{sukhbold:2014} were employed, though for the referenced progenitor masses the models are very similar between the two sets.}
While it is possible that the artificially imposed axisymmetry and the concomitant incorrect dynamics as compared with 3D simulations \citep{hanke:2012, couch:2013a, couch:2014, dolence:2013} are to blame for this difference, more likely is that these 1D models for artificially driving the explosions are missing some important aspect of the CCSN mechanism. 

An obvious candidate for a key piece of CCSN physics that is missing from previous models for 1D explosions is {\it turbulence}. 
A number of recent works, many based on 3D simulations, have pointed out the key role that turbulence plays in the CCSN mechanism \citep{murphy:2011, murphy:2013, hanke:2012, couch:2013a, couch:2015a, radice:2016, radice:2018a, mabanta:2018}.
Turbulence, through the chaotic motion of eddies, provides an effective pressure that supports shock expansion \citep{murphy:2013, couch:2015a}, plays a key role in the transport of energy and composition \citep{radice:2016}, and results in the significant dissipation of kinetic energy to heat \citep{mabanta:2018}. 
Using parameterized neutrino leakage simulations in 1D, 2D, and 3D, \citet{couch:2015a} show that the amount of neutrino energy absorbed in the CCSN gain layer is not that different between a successful 3D explosions and a failed 1D explosion.
Indeed, in order to drive neutrino-driven explosions in 1D simulations required artificially enhancing the neutrino heating to levels far beyond what was observed in 3D explosions.
The difference, \citet{couch:2015a} argue, is made up by the action of turbulence in aiding shock expansion.
This raises concerns about the accuracy of 1D parameterizations that rely on enhancing neutrino heating to drive explosions. 

In this article, we present a new parameterized method for driving CCSN explosions in 1D that includes the most salient features of convection and turbulence. 
We call our new approach Supernova Turbulence In Reduced-dimensionality, or {\it STIR}.
Inspired by the works of \citet{murphy:2011, murphy:2013}; and \citet{mabanta:2018}, we begin with a Reynolds decomposition of the fluid equations that separates the flow variables into background, {\it mean} components and perturbed, turbulent components. 
Extending these previous works, we use the fully time-dependent, non-steady-state forms of the equations. 
We then {\it angle-average} the full Reynolds-decomposed equations, reducing them to a set of 1D evolution equations. 
After making certain simplifying assumptions appropriate for the CCSN context, the equations include terms that depend essentially on a single turbulent parameter: the characteristic speed of turbulent eddies. 
Since CCSN turbulence is driven primarily by convection, we use a modified, time-dependent version of mixing-length theory (MLT) to estimate the evolution of this typical turbulent speed. 
We compare this model to full, high-fidelity 3D simulations of the CCSN mechanism taken from \citet{oconnor:2018b}.
We find that STIR is able to reproduce the strength and locality of turbulent motions, in an angle-averaged sense, extremely well, and is also able to better model gross features of the dynamics such as the evolution of the shock far better than 1D simulations that neglect turbulence.

STIR makes no ad hoc modifications to the neutrino transport or microphysics of the CCSN simulations. 
We include full, two-moment, energy-dependent neutrino transport \citep{oconnor:2015}, precisely as we use in multidimensional simulations \citep{oconnor:2018, oconnor:2018b}, without any modifications to, e.g., the neutrino interactions, cross sections, or heating rates.
We include the full PNS and do not excise any portion of the inner core, allowing us to directly explore the sensitivity of our 1D CCSN simulations to, e.g., the nuclear EOS and other nuclear physics properties of the PNS.
The small number of free parameters that enter our model are chosen on the basis of comparison to full 3D simulations of the CCSN mechanism, and not chosen in order to reproduce any particular observed feature of CCSNe.
In general, we find that STIR reproduces the features of multidimensional CCSN simulations quite well, including which stars explode and which fail, resulting in collapse of the PNS to a black hole (BH).
We perform a parameter study with STIR in solar metallicity progenitor stars from 9 \Msun to 120 \Msun. 
We find reasonable agreement with observed statistics of the CCSN population such as explosion fraction and remnant masses distributions. 

Very recently, \citet{mabanta:2019} have presented a model for driving CCSN explosions in 1D including the effects of turbulent convection that is similar in many respects to STIR.
They also start from a Reynolds decomposition of the flow variables, but their final model is distinct from ours in a number of key ways.
We briefly compare their model and STIR.

This paper is organized as follows.
In Section \ref{s.turb} we present the derivation of the STIR model and discuss our use of MLT as a closure.
In Section \ref{s.num} we discuss our numerical implementation and inclusion of STIR in our CCSN mechanism code.
We compare the results of STIR simulations for a 20-\Msun progenitor to the 3D simulations of \citet{oconnor:2018b} in Section \ref{s.stir3d}. 
In Section \ref{s.param} we present a first parameter study using STIR for the same progenitor set employed by \citet{sukhbold:2016} and compare our results to theirs, and other similar parameter studies.
We conclude and discuss the future outlook for this new model for 1D CCSN explosions in Section \ref{s.conc}.

\section{Turbulent convection in 1D}\label{s.turb}
\subsection{Turbulent correlation terms}

The presence of turbulence and convection changes the dynamics of stars and supernovae fundamentally.
In this section, we describe the salient equations governing these dynamics and our model for incorporating turbulent convection into 1D CCSN simulations.
Our approach is related to that of \citet{bruenn:1995} and, more distantly, \citet{wilson:1988} and \citet{bohm-vitense:1958}, though modified significantly.
We have drawn inspiration for our approach from the work of, e.g., \citet{meakin:2007, murphy:2011, murphy:2013, arnett:2015, mabanta:2018}.

The compressible Euler equations describing the conservation of mass, momentum, and energy for a self-gravitating system are
\begin{align}
  \partial_t \rho + \nabla \cdot (\rho \bm{u}) &= 0,\label{e.mass}\\
  \partial_t (\rho \bm{u}) + \nabla \cdot (\rho \bm{u} \otimes \bm{u} + P\mathbf{I}) &= -\rho \bm{g},\label{e.mom}\\
  \partial_t (\rho e) + \nabla \cdot [\bm{u} (\rho e + P)] &= -\rho \bm{u}\cdot \bm{g} \label{e.ener},
\end{align}
where $\rho$ is the mass density, $\bm{u}$ is the velocity vector, $P$ is the pressure, $e=e_i + \tfrac{1}{2}u^2$ is the total specific energy, $e_{i}$ is the internal energy, $\bm{g}$ is the gravitational acceleration, and $\mathbf{I}$ is the identity tensor.
The impact of turbulence on the dynamics of compressible flows can be modeled by decomposing the flow variables into a background, {\it mean} component and a perturbed, or turbulent, component: $\phi = \phi_0 + \phi^\prime$.
By definition, $\langle \phi \rangle = \phi_0$, where $\langle ... \rangle$ represents a suitable averaging in space and time, requiring $\langle \phi^\prime \rangle = 0$.
For instance, the velocity vector is the sum of its mean and turbulent components, $u_i = v_i + v_i^\prime$, where $v_i = \langle u_i \rangle$.
Applying such a decomposition and averaging procedure to Equations (\ref{e.mass})-(\ref{e.ener}), a so-called Reynolds averaging, yields additional terms related entirely to the turbulent character of the flow.
In the context of CCNSe \citep{murphy:2011}, the three most significant turbulent correlation terms are the Reynolds stress tensor $\mathbf{R}$, the energy flux due to turbulence $\bm{F}_e$, and the turbulent dissipation $\epsilon_\mathrm{turb}$ \citep{mabanta:2018}.

The modified, {\it turbulent} Euler equations, including only the most significant turbulent correlations, are then
\begin{align}
  \partial_t \langle \rho \rangle + \nabla \cdot (\rho_0 \bm{u}_0) = &\ 0,\label{e.tmass}\\
  \partial_t \langle \rho \bm{u} \rangle + \nabla \cdot (\rho_0 \bm{u}_0 \otimes \bm{u}_0 + P_0 \mathbf{I}) = &-\rho_0 \bm{g} \nonumber\\
  &- \nabla \cdot \langle \rho \mathbf{R} \rangle,\label{e.tmom}\\
  \partial_t \langle \rho e \rangle + \nabla \cdot [\bm{u}_0 (\rho_0 e_0 + P_0)] = &-\rho_0 \bm{u}_0\cdot \bm{g} \label{e.tener} \nonumber\\
  &- \nabla \cdot \bm{u}_0 \langle \rho \mathbf{R} \rangle \nonumber\\
  &- \nabla \cdot \bm{F}_e \nonumber\\
  &+ \rho_0 \epsilon_\mathrm{turb}.
\end{align}
The above equations neglect certain higher-order turbulent correlation terms that may be important in certain regimes.
Specifically, we ignore any turbulence-induced pressure perturbations, $P^\prime$, and attendant terms (i.e., the Boussinesq approximation).
This approximation is only valid for low turbulent Mach numbers, which is generally fine for most regimes of stellar convection, but can become a poor assumption during the onset of explosion in CCSNe (see \citet{murphy:2013} and \citet{couch:2015a}).

These equations are supplemented by an evolution equation of the specific turbulent kinetic energy, $K$, which is defined as one-half of the trace of the Reynolds stress: $K = \tfrac{1}{2} \mathrm{Tr}(\mathbf{R})$.
The full turbulent kinetic energy equation is \citep{murphy:2011, mabanta:2018},
\begin{eqnarray}
  \frac{\partial \langle \rho K \rangle}{\partial t} + \nabla \cdot (\langle \rho K \rangle \bm{v}) = &-& \mathrm{Tr}(\langle \rho \mathbf{R} \rangle \cdot \nabla\bm{v}) + \langle \rho^\prime \bm{v}^\prime \rangle
  \cdot \bm{g} \nonumber \\
  &-& \nabla \cdot \langle \bm{F}_K \rangle - \nabla \cdot \langle \bm{F}_P \rangle \nonumber\\
  &+& \langle P^\prime \nabla \cdot \bm{v}^\prime \rangle - \rho_0 \epsilon_\mathrm{turb},
\end{eqnarray}
where $\bm{F}_K = \rho K \bm{v}^\prime$ is the turbulent kinetic energy flux and $\bm{F}_P = P^\prime \bm{v}^\prime$ is the turbulent pressure flux.
The trace term (first term on the RHS) is the production of turbulence due to shear. 
If we again assume that the pressure fluctuation induced by turbulence is negligible then the turbulent energy equation in spherical symmetry becomes,
\begin{align}
  \frac{\partial \langle \rho K \rangle}{\partial t} + \frac{1}{r^2}\frac{\partial}{\partial r} [r^2 (\langle \rho K \rangle v_r + \langle \rho K v_r^\prime \rangle)]
  = &- \langle \rho R_{rr} \rangle \frac{\partial v_r}{\partial r} \nonumber\\
    &+ \langle \rho^\prime v_r^\prime \rangle g \nonumber\\
    &- \rho \epsilon_\mathrm{turb}. \label{e.evolK}
\end{align}
So, we shall assume that turbulent energy is generated by shear and buoyancy (first and second terms on RHS), destroyed by dissipation (third term on RHS), advected with the background flow (first term in divergence on LHS), and diffused (second term in divergence on LHS).
In the shear term, we have assumed spherical symmetry and neglected the possibility of background rotational flow. 
The presence of background rotation would lead to additional shear terms. 
For the buoyancy term, we have assumed that the gravitational acceleration, $g$, is purely radial.

The rate of dissipation of turbulent kinetic energy to heat is
\begin{equation*}
  \epsilon_\mathrm{turb} = \mathrm{Tr} (2 \nu (\nabla \bm{v}^\prime) \cdot (\nabla \bm{v}^\prime))/2,
\end{equation*}
where $\nu$ is the fluid viscosity.
Following \citet{kolmogorov:1941} and \citet{mabanta:2018}, we can relate the turbulent dissipation to the Reynolds stress, so for spherical symmetry
\begin{equation}
  \epsilon_\mathrm{turb} \approx \frac{R_{rr}^{3/2}}{\Lambda} = \frac{v_r^{\prime 3}}{\Lambda},\label{e.epsturb}
\end{equation}
where $\Lambda$ is the largest scale on which turbulent energy is dissipated, i.e., the largest turbulent eddy size.

Turbulence appears in the Reynolds-averaged momentum equation (Equation~\ref{e.tmom}) via the Reynolds stress,
\begin{equation}
  R_{ij} = v_i^\prime v_j^\prime,
\end{equation}
which yields a source term  of the form $-\nabla \cdot \langle \rho \mathbf{R} \rangle$.
Owing to the fact that the turbulence is driven by buoyant convection, the Reynolds stress in 3D simulations of CCSNe is anisotropic \citep{murphy:2013, couch:2015a, radice:2016}.
Thus, in a spherical coordinate basis, the radial-radial component is roughly equally to the sum of the transverse diagonal components,
\begin{equation}
  R_{rr} \sim R_{\theta \theta} + R_{\phi \phi},\label{e.anisor}
\end{equation}
and the transverse diagonal components are approximately equal, $R_{\theta \theta} \sim R_{\phi \phi}$.
The trace of the Reynolds stress is then $\mathrm{Tr}(\mathbf{R})\approx2R_{rr}$.
In spherical symmetry, therefore, we can simplify the turbulent momentum source term:
\begin{equation}
  -\nabla \cdot \langle \rho \mathbf{R} \rangle \approx -\frac{1}{r^2} \frac{\partial}{\partial r} (r^2 \rho v_r^{\prime 2}). \label{e.divpturb}
\end{equation}
The quantity $\rho v_r^{\prime 2}$ has units of pressure and so is often defined as the {\it turbulent pressure}, $P_\mathrm{turb}$.

Turbulent stresses transport heat in the fluid. This heat transport results in an additional source term in the energy equation:
\begin{equation*}
  -\nabla \cdot \bm{F}_e = - \nabla \cdot  \langle \rho \bm{v}^\prime e^\prime \rangle
\end{equation*}
where $\bm{F}_e$ the internal energy flux due to turbulence and $e^\prime$ is the internal energy fluctuation. Since $R_{\theta \theta} \sim R_{\phi \phi}$, there will be no net transport of heat via turbulence in the transverse directions, so in spherical symmetry
\begin{equation}
  -\nabla \cdot \bm{F}_e = -\frac{1}{r^2} \frac{\partial}{\partial r} (r^2 \rho v_r^\prime e^\prime ). \label{e.divfturb}
\end{equation}

\subsection{Mixing length theory closure}

Following the assumption of spherical symmetry and anisotropic turbulent stresses obeying Equation (\ref{e.anisor}), we arrive at expressions for the turbulent correlations (Equations (\ref{e.epsturb}), (\ref{e.divpturb}),  (\ref{e.divfturb})) that depend on only a single turbulent quantity, the characteristic turbulent speed in the radial direction, $v_r^\prime$.
This, in turn, evolves according to Equation (\ref{e.evolK}).
Now we must find a means to relate the turbulent speed to other unknowns of the system in order to solve for its evolution; in other words, we need a {\it closure} for our turbulence model.
For this, we appeal to mixing length theory \citep[MLT;][]{bohm-vitense:1958, cox:1968}, assuming that the turbulence in CCSNe is driven by convection.
This is an incomplete picture as other instabilities, most notably the standing accretion shock instability (SASI), can also drive turbulence \citep{endeve:2010, endeve:2012}.
Nevertheless, we shall make the approximation that the typical turbulent speed is equivalent to the typical convective speed, $\vtrb \equiv v_r^\prime \sim v_\mathrm{con,MLT}$, the latter computed via a modified MLT described below.

MLT relates the transport of energy and compositional mixing to the typical speed of a putative buoyant blob rising against the background flow.
Such a buoyant blob will experience an acceleration described by the local \brunt frequency.
Simultaneously, a buoyant blob that begins to rise against the background flow will experience a drag force, resulting in turbulent dissipation of the blob's kinetic energy.
Following MLT, the buoyant forcing is
\begin{equation}
  \langle \rho^\prime v_r^\prime \rangle g \approx \rho \vtrb \omega_\mathrm{BV}^2 \Lambda_\mathrm{mix},
\end{equation}
where $\omega_\mathrm{BV}$ is the \brunt frequency and $\Lambda_\mathrm{mix}$ is the mixing length.
We calculate the \brunt frequency assuming the Ledoux criterion for convection,
\begin{equation}
  \omega_\mathrm{BV}^2 = -g_\mathrm{eff} \left (\frac{1}{\rho}\frac{\partial \rho}{\partial r} - \frac{1}{\rho c_s^2}\frac{\partial P}{\partial r} \right ), \label{e.fbv2}
\end{equation}
where $c_s$ is the adiabatic sound speed.
This expression for the Ledoux \brunt frequency is completely equivalent to expressions that explicitly include entropy and electron fraction gradients \citep{muller:2016a}.
We have experimented with several other expressions for the \brunt frequency, including those using entropy and lepton gradients.
We find that using Equation (\ref{e.fbv2}), which avoids the need to compute thermodynamic derivatives and any additional spatial gradients, generally results in the smoothest $\omega_\mathrm{BV}^2$.
Our sign convention is such that positive $\omega_\mathrm{BV}^2$ implies convective instability.

In computing the \brunt frequency, we modify the gravitational acceleration to account for local acceleration from the background flow, yielding an effective gravitational acceleration,
\begin{equation}
    g_\mathrm{eff} = -\frac{\partial \Phi}{\partial r} + v_r \frac{\partial v_r}{\partial r},
    \label{e.geff}
\end{equation}
where $\Phi$ is the gravitational potential. 
In effect, this term shifts the frame of reference in which the gravitational acceleration is computed. 
During the early post-bounce, pre-explosion phase this modification to the acceleration is negligible. 
Only once an explosion begins are $v_r$ or its gradient very large. 
Following explosion, throughout most of the ejecta both the background radial velocity and its gradient are positive. 
Thus, the effect of including the second term on the RHS of Equation (\ref{e.geff}) is to reduce the magnitude of the gravitational acceleration felt in the ejecta. 
This drives the \brunt closer to zero, shutting off buoyant acceleration in regions that are exploding. 
Physically, this reflects that the ``buoyant plumes'' of the convection have essentially become the background flow once an explosion sets in. 

In the PNS, the enhanced lepton fraction gradient induced by trapped neutrinos can also drive convection \citep{wilson:1988}.
This effect is neglected in Equation (\ref{e.fbv2}).
Including the total lepton fraction gradient correctly requires complicated thermodynamic derivatives in the \brunt frequency \citep{roberts:2012a} and we find that the overall effect is small in the gain region, which is what we are most concerned about in this work.  
We leave an improved treatment of the PNS convection to future work.

For the mixing length, we take a fraction of the pressure scale height as computed from the equation of hydrostatic equilibrium,
\begin{equation}
  \Lamm= \alphL H_P = \alphL \frac{P}{\rho g},
  \label{e.Lamm}
\end{equation}
where $g = - \partial\Phi / \partial r$ is the magnitude of the local gravitational acceleration and $\alphL$ is a tunable parameter.  This is a standard approximation to the mixing length in MLT.
Of course hydrostatic equilibrium is not a very good assumption for a collapsing stellar core, but it is not so bad for the post-shock region. 
And, critically, we find that direct calculation of the pressure scale height via the gradient of the pressure introduces unwanted oscillations near the shock.
The pressure scale height diverges at the coordinate origin due to the $g^{-1}$ term, so we limit the mixing length to be no larger than the local radial coordinate.
In practice, this only occurs deep inside the PNS, below any convective regions.

Following the standard assumption in MLT, we relate the diffusive flux due to turbulent convection of some scalar $X$ to its local gradient via \citep[cf.][]{cox:1968, wilson:1988}
\begin{equation}
  v_r^\prime X^\prime \approx - \alpha_X \vtrb \Lamm \nabla X \equiv - D_X \nabla X,
\end{equation}
where we have defined the diffusion coefficient $D_X$.
Thus, the diffusive flux of turbulent energy is
\begin{equation}
  \langle \rho K v_r^\prime \rangle \approx -\rho D_K \nabla K,
\end{equation}
where the corresponding diffusion coefficient is
\begin{equation}
  D_K = \alpha_K \vtrb \Lambda_\mathrm{mix}.\label{e.Dk}
\end{equation}
$\alpha_K$ is a tunable parameter to control the rate of diffusion.
The dissipation/drag term is simply Equation (\ref{e.epsturb}), where we use the mixing length for the dissipation scale.
So the turbulent energy equation, Equation (\ref{e.evolK}), then becomes,
\begin{align}
  \frac{\partial(\rho \vtrb^2)}{\partial t} &+ \frac{1}{r^2}\frac{\partial}{\partial r} [r^2 (\rho \vtrb^2 v_r - \rho D_K \nabla \vtrb^2)]\nonumber\\
  &= - \rho \vtrb^2 \frac{\partial v_r}{\partial r} + \rho \vtrb \omega_\mathrm{BV}^2 \Lamm - \rho \frac{\vtrb^3}{\Lambda_\mathrm{mix}},\label{e.eom}
\end{align}
where we have used $R_{rr} \sim \vtrb^2$ in the shear term (first term on RHS).
In general, the shear is quite a bit smaller than either buoyancy or turbulent dissipation. 
The last term on the RHS of Equation (\ref{e.eom}) is the rate of dissipation of turbulent energy to heat, Equation (\ref{e.epsturb}), where we have assumed that the dissipation scale is equivalent with the mixing length. 
This is not necessarily case as the physics governing these two scales are, in principle, different. 
This assumption, however, reduces the free parameters required by the model and results in peak convective speeds that agree with the expectations from MLT.

The peak speed of the convection can be found by setting the LHS of Equation (\ref{e.eom}) to zero, assuming a steady state with no background velocity gradient, and solving for speed,
\begin{equation}
  v_\mathrm{turb,max} = \omega_\mathrm{BV} \Lambda_\mathrm{mix}, \label{e.vcmax}
\end{equation}
which is identical to the usual expression for the convective speed in standard MLT implementations \citep{cox:1968, paxton:2013, muller:2016a}.
Rather than assume that the convection becomes instantaneously fully-developed with the typical speed given by Equation (\ref{e.vcmax}), we solve Equation (\ref{e.eom}) for the time-dependent local convective speed.
We allow for negative squared \brunt frequencies, i.e., negative buoyancy, resulting in deceleration of convective flow in regions that are stably stratified. 
This is particularly important in the neutrino cooling region below the gain region. 
There, convective plumes falling down from the gain layer are rapidly decelerated by the strongly positive entropy and lepton gradients. 

We integrate Equation (\ref{e.eom}) in an operator split fashion.
During the hydrodynamic update, the turbulent energy is advected with the flow and the hyperbolic fluxes are modified by the diffusive turbulent flux assuming an exact conservation law (i.e., the RHS of Equation (\ref{e.eom}) is set to zero).
The buoyant and dissipative source terms in Equation (\ref{e.eom}) are incorporated separately from the hydrodynamic update using a simple forward-Euler approach.
Since some small perturbation is required to seed convection, we assume that the minimum $\vtrb$ in regions of positive $\omega_\mathrm{BV}^2$ is
\begin{equation}
  v_\mathrm{turb,min} = \Delta t\ \omega_\mathrm{BV}^2 \Lambda_\mathrm{mix},
  \label{e.vtrbmin}
\end{equation}
where $\Delta t$ is the computational time step size.
As a consequence of advecting the turbulent energy with the flow, in regions of sufficient background radial velocity, $\vtrb$ will be advected out of layers with positive $\omega_\mathrm{BV}^2$ and will be damped in time according to the turbulent dissipation and buoyant deceleration.
This reflects the requirement of sufficiently rapid growth of convection (i.e., sufficiently large $\omega_\mathrm{BV}^2$) for convection to become strong in the presence of a background accretion flow \citep{foglizzo:2006}.

A time-dependent treatment of the convective speed is justified by a consideration of relevant time scales in the problem.
Assuming the background is stationary ($v_r\sim 0$), we can integrate Equation (\ref{e.eom}) from $\vtrb=0$ to some fraction $f$ of $v_\mathrm{turb,max}$ to find the growth time of the convection,
\begin{equation}
  \tau_\mathrm{con} = \frac{\tanh^{-1}(f)}{\omega_\mathrm{BV}}.\label{e.tauCon}
\end{equation}
Thus, the true $v_\mathrm{turb,max}$ is only reached asymptotically at infinite time.
A typical $\omega_\mathrm{BV}^2$ in the gain region of a CCSN is $\sim 10^5$ s$^{-2}$, thus the time for the convective speed to reach 90\% of its maximum is $\sim$10 ms.
This is remarkably similar to the advection time through the gain region, $\tau_\mathrm{adv} = \Delta r_\mathrm{gain} / \langle v \rangle_\mathrm{gain} \sim 50\ \mathrm{km}/5000\ \mathrm{km\ s^{-1}} \sim 10\ \mathrm{ms}$.
This is also roughly the dynamical time scale, $(\rho G)^{-1/2}$, at the shock radius.
All of this indicates that the growth of convection occurs on time scales similar to other processes in the CCSN gain layer, i.e., it is not {\it fast} and should not be treated as instantaneous.
This has also been shown in multidimensional simulations wherein we observe convection developing ``slowly'' and from the analysis of \citet{foglizzo:2006}.
Integrating Equation (\ref{e.eom}) with $\omega_\mathrm{BV}^2=0$ we find
the characteristic time scale for the convection to slow down from this peak speed in stable regions: $\tau_\mathrm{drag} = \Lamm\vtrb^{-1} \sim 50\ \mathrm{km} / 2000\ \mathrm{km\ s^{-1}} \sim 25\ \mathrm{ms}$.
Again, a comparable time scale to others in the problem, if slightly slower.

Finally, it is worth noting that we have essentially equated the convective speed to the square root of the turbulent kinetic energy. 
For Kolmogorov-like turbulence, as is the case in the CCSN gain region \citep{radice:2016}, the largest turbulent scales contain the vast majority of the kinetic energy. 
Our equating of the average convective speed to the turbulent speed is consistent with this characteristic of CCSN turbulence.

\subsection{Modified evolution equations}

We are now equipped to compute the turbulent correlation terms, Equations (\ref{e.epsturb})-(\ref{e.divfturb}), in a space- and time-dependent fashion. 
With the model for turbulence and convection in the CCSN context described above, the resulting evolution equations for mass, momentum, energy, electron fraction, and turbulent kinetic energy are
\begin{align}
  \frac{\partial \rho}{\partial t} &+ \frac{1}{r^2}\frac{\partial}{\partial r}\left[r^2 \rho v_r \right] = 0,\label{e.mass3}\\
  \frac{\partial (\rho v_r)}{\partial t} &+ \frac{1}{r^2}\frac{\partial}{\partial r}[r^2(\rho v_r^2 + P + \rho \vtrb^2 )] \nonumber\\
  &= -\rho g + \mathcal{S}_\nu, \label{e.mom3} \\ 
  \frac{\partial(\rho e)}{\partial t} &+ \frac{1}{r^2}\frac{\partial}{\partial r}[r^2 v_r(\rho e + P + \rho\vtrb^2 - \rho D_e\nabla e)] \nonumber\\
  &= -\rho v_r g + \rho\frac{\vtrb^3}{\Lamm} + \mathcal{Q}_\nu, \label{e.ener3} \\ 
  \frac{\partial(\rho Y_e)}{\partial t} &+ \frac{1}{r^2}\frac{\partial}{\partial r}[r^2 v_r (\rho Y_e - \rho D_{Y_e}\nabla Y_e)] = \mathcal{C}_\nu,\label{e.ye} \\ 
  \frac{\partial(\rho \vtrb^2)}{\partial t} &+ \frac{1}{r^2}\frac{\partial}{\partial r} [r^2 (\rho \vtrb^2 v_r - \rho D_K \nabla \vtrb^2)]\nonumber\\
  &= - \rho \vtrb^2 \frac{\partial v_r}{\partial r} + \rho \vtrb \omega_\mathrm{BV}^2 \Lamm- \rho \frac{\vtrb^3}{\Lambda_\mathrm{mix}}, \label{e.turbk}
\end{align}
where $\mathcal{S}_\nu$, $Q_\nu$, and $\mathcal{C}_\nu$ are source terms due to matter-neutrino interactions and $g$ is the gravitational acceleration. In Equations (\ref{e.ener3})-(\ref{e.turbk}), the respective diffusion coefficients are 
\begin{eqnarray}
  D_e &=& \alpha_e \vtrb \Lamm, \\
  D_{Y_e} &=&
   \alpha_{Y_e} \vtrb \Lamm, \\ \label{e.diffYe}
  D_{K} &=& \alpha_K \vtrb \Lamm.
\end{eqnarray} 
In general, the various diffusion parameter $\alpha$'s can have different, independent values but for the present work we assume they are all equal and $\alpha_K = \alpha_e = \alpha_{Y_e} = 1/6$. 

Equations (\ref{e.mass3})-(\ref{e.turbk}) describe the dynamics of a one-dimensional system in the presence of non-spherical, turbulent motion and neutrino radiation. 
It is perhaps a subtle, even semantical, point to make but this is no longer truly a {\it spherically-symmetric} system, owing to the inclusion of turbulent, convective motion.
More precisely, this might now be described as an {\it angle-averaged} approach to the full dynamics.
It is also worth commenting on the conservation of energy.
As pointed out by \citet{mabanta:2018}, the energy in turbulent convection in the CCSN context is extracted from the free energy in unstable thermodynamic and compositional gradients.
This is accounted for in our model described above. 
The turbulent kinetic energy is generated by buoyancy (Equation (\ref{e.eom})). 
This buoyancy, in turn, is the product of unstable gradients, as described by the \brunt frequency (Equation (\ref{e.fbv2})). 
The diffusive mixing induced by the turbulent convection flattens these gradients, reducing the buoyant driving.
In the limit of fully efficient convection, the gradients will be eliminated along with the buoyant driving.
Thus, up to a factor of order unity (determined by the diffusive mixing $\alpha$ parameters), total energy is conserved by STIR, when accounting also for the free energy in unstable thermodynamic and compositional gradients.

Our approach for including turbulent convection in 1D CCSN simulations recalls that of \citet{bruenn:1995}.
There, the authors include MLT for the diffusive convective transport of energy, composition, and neutrinos. 
They also include a turbulent pressure term in the momentum equation. 
Our approach extends this is a number of ways.
First, \cite{bruenn:1995} assume that the turbulence is isotropic, whereas we account for the fact that in the CCSN gain region is, in fact, quite anisotropic \citep[Equation (\ref{e.anisor}),][]{couch:2015a}. 
We also account for many more turbulent correlation terms in the evolution equations than just the pressure term in the momentum equation.
These terms are key to accurately modeling realistic 3D CCSN convection. 

\citet{yamasaki:2006} also explore the effects of convection in 1D CCSN models. 
This work relied on steady-state models and a phenomenological approach to including the effects of convection.
As in \citet{bruenn:1995}, they neglected several terms related to the presence of turbulence that we now understand to be critical. 
The model of \citet{yamasaki:2006} primarily included the effect of diffusive energy transport outward toward the shock as an aid to shock expansion. 
This effect is very important, but only part of the story of the impact of turbulence in CCSNe. 

STIR is most reminiscent of the recent work of \citet{mabanta:2018}. 
They include all the same turbulent correlation terms we do but in the context of a steady-state system and using parameterized source terms to treat the neutrino physics. 
They were the first to point out the key role of the turbulent dissipation term and our experiments with STIR confirm this.
The turbulent dissipation term is at least as important as the ``turbulent pressure'' terms appearing in the momentum and energy equations. 
In contrast to \citet{mabanta:2018}, in STIR we treat the turbulent convection terms in a fully time- and space-dependent manner, adopting a new closure based on MLT. 
We also include high-fidelity, energy-dependent neutrino transport in our model.

\citet{mabanta:2019} extend the model of \citet{mabanta:2018} to time-dependent 1D simulations.
Their approach includes most of the same turbulent correlation terms we include in STIR, though their approach to closing the model is completely different.
Whereas here we have used a time-dependent MLT approach, \citet{mabanta:2019} relate the strength of turbulence and convection directly to the neutrino luminosity \citep{murphy:2011, murphy:2013}. 
Furthermore, they treat neutrinos with a simple heating/cooling ``lightbulb'' approach with a constant luminosity that is input by-hand \citep{murphy:2008}.
This requires them to make assumptions about the radial dependence of the convective terms through the gain region, since these quantities are local in nature. 
In STIR, our use of MLT relates the convective terms to the local thermodynamic gradients and, as such, is more general, allowing convection to be driven by other physical mechanisms besides just neutrino heating. 
We also include high-fidelity, energy-dependent neutrino transport and approximate general relativistic gravity (Section \ref{s.num}). 
\citet{mabanta:2019} assume purely Newtonian gravity.

\section{Numerical Approach}
\label{s.num}

Our CCSN application is implemented in the FLASH adaptive mesh refinement (AMR) simulation framework \citep{fryxell:2000, dubey:2009}.
We solve Equations (\ref{e.mass3})-(\ref{e.turbk}) using a newly-implemented hydrodynamics solver based on a fifth-order finite-volume weighted essentially non-oscillatory (WENO) spatial discretization \citep{shu:1988, tchekhovskoy:2007, shu:2009} and a method-of-lines Runge-Kutta time integration.
We use the WENO steepness indicators of \citet{borges:2008} and a two-stage second-order SSP Runge-Kutta time integrator \citep[e.g.,][]{shu:1988}.
Details of this new solver will be presented in a forthcoming methods paper \citep{couch:2019a}.
We use an HLLC Riemann solver everywhere except in shocks, where we use a more diffusive HLLE solver \citep{toro:2009}.
We treat self-gravity using an approximate general relativistic effective potential \citep{marek:2006, oconnor:2018}.
We use the ``optimal'' EOS parameterization of \citet[][hereafter SFHo]{steiner:2013}.
In the present work, we assume nuclear statistical equilibrium (NSE) abundances everywhere. 

We include the additional turbulent correlation terms on the RHS of Equations (\ref{e.mass3})-(\ref{e.turbk}), including the diffusive mixing terms, directly in the explicit hyperbolic fluxes. 
We monitor the parabolic diffusive time step limit to ensure stability, though since our neutrino transport scheme is also explicit, the parabolic time step is almost always larger than the transport time step.

For the transport of neutrinos, we employ an explicit two-moment scheme with analytic closure for higher moments \citep[so-called ``M1'' transport,][]{shibata:2011, cardall:2013, oconnor:2015}.
Our implementation of M1 transport in FLASH is detailed in \citet{oconnor:2018}.
In the present study, we use 12 energy groups spaced logarithmically.
Our base opacity set from NuLib \citep{oconnor:2015} closely matches that of \citet{bruenn:1985}, with corrections for weak magnetism following \citet{horowitz:2002}.
We include velocity-dependent transport terms and inelastic neutrino-electron scattering according to \citet{oconnor:2015}.

We make no ad-hoc modifications to the neutrino physics in our simulations except to include turbulent diffusion of trapped neutrinos.
Analogous to the diffusive mixing terms for advected mass scalars such as the electron fraction (Equation (\ref{e.ye})), we include an additional term in the explicit hyperbolic fluxes of neutrino energy density \citep{wilson:1988,bruenn:1995},
\begin{equation}
  \mathcal{F}_{\nu,\mathrm{dif}} = \left( D_\nu \nabla E_{\nu} \right)(1-f)^4
\end{equation}
where $E_{\nu}$ is the energy density of the neutrino radiation field (the zeroth moment of the neutrino distribution function in our M1 scheme).
The neutrino flux factor, $f = F_\nu / E_\nu$, limits to 1 in regions where the neutrinos are free-streaming and to $\sim$0 in diffusive regions.
The $(1-f)^4$ term, then, smoothly shuts off the diffusive mixing of neutrinos in low optical depth layers where neutrinos are not trapped.
Similar to the other diffused scalar fields, the diffusion coefficient for neutrino energy density is 
\begin{equation}
  D_\nu = \alpha_\nu \vtrb \Lamm, \label{e.diffNu}
\end{equation}
and in the present working we assume $\alpha_\nu = 1/6$.

For all of the simulations described here, we use 10 levels of refinement in a domain with radial of 15,000 km, yielding a finest grid spacing of 0.244 km. 
We limit the maximum allowed level of refinement logarithmically with radius with a typical $\Delta r/r$ of 0.7\%.

\section{Comparison to 3D CCSN turbulence} \label{s.stir3d}

We now compare our 1D model for turbulent convection in the CCSN context to fully 3D simulations. 
For this, we use the 3D data from \citet{oconnor:2018b}, their model {\tt mesa20\_LR\_v} which includes velocity dependence in the neutrino transport and inelastic scattering on electrons up to 16 ms post-bounce. 
The progenitor is a 20-\Msun model from \citet{farmer:2016}.
We construct radial profiles of various quantities from the 3D data by angle averaging the full 3D data at 135 ms post-bounce.
This is just after the maximum shock extension prior to subsequent shock recession, a time when convection is fully developed and strong in both the gain region and the PNS.
In comparing to 3D, we vary only the mixing length parameter, $\alphL$, and keep all other free parameters of the STIR model fixed.
In principle, fully fitting to the 3D data would require we allow all the model parameters to vary simultaneously, which will be explored in future work.

\begin{figure}[tb]
  \centering
  \includegraphics[width = 0.47\textwidth]{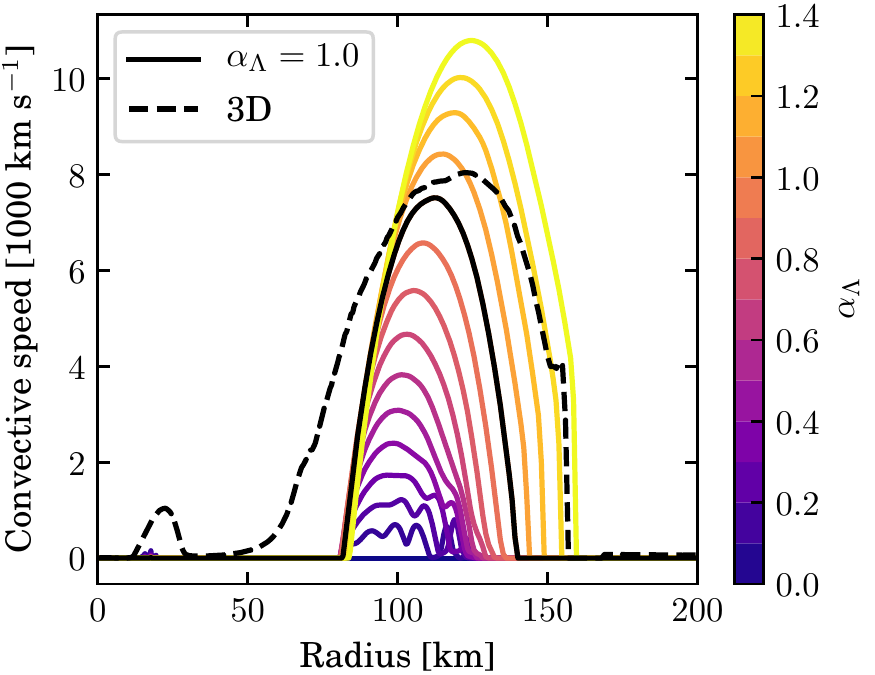}
  \caption{
    Angle-averaged turbulent speed \vtrb from the 3D simulation of \citet{oconnor:2018b} (dashed line) along with the turbulent speed from STIR simulations with a range of $\alphL$ values spaced in intervals of 0.1.  All simulations are at 135 ms post-bounce. 
    $\alphL=1.0$ is marked with heavy black line to guide the eye.
    Values of $\alphL$ around 1.1 yield similar peak convective speeds to the 3D simulation while larger values, between 1.2 and 1.3, reproduce the total amount of convective kinetic energy.
    The more extended ``tail'' of convection below the gain layer seen in the 3D simulation between 50 km and 80 km is a product of angular variation in the 3D model and not of excessive diffusion or advection of convective motion along any given line of sight in 3D as compared to STIR.
    STIR does not reproduce the PNS convection around 25 km very well.
  }
  \label{f.alpha_vcon}
\end{figure}

In Figure \ref{f.alpha_vcon} we show the turbulent speed \vtrb from the 3D simulation along with the same from our STIR simulations for various values of $\alphL$.
For the 3D data, we define the turbulent speed based on the Reynolds stress tensor as in \citet{couch:2015a}, analogous to our definition in Section \ref{s.turb}.
STIR results in turbulent velocity profiles that are very similar to those from the 3D simulation in both location and strength. 
Figure \ref{f.alpha_vcon} implies that the ``best fit'' value of \alphL is somewhere between 1.2 and 1.3 when considering the integrated amount of turbulent kinetic energy, while \alphL values around 1.1 yield the closest match to the peak speeds. 
And, as expected, the strength of the turbulent convection in STIR is a strong function of the mixing length parameter $\alphL$. 
A notable exception to this is for convection in PNS, which is clearly evident in the 3D simulation around 25 km in radius.
STIR predicts essentially no convection here.
Adjusting the various $\alpha$ parameters in our model associated with the strength of diffusive mixing can yield a better match to the PNS convection in 3D, but for the sake of simplicity we do not explore this in the present work and leave it to future work. 
Also, we are mostly concerned with the turbulent convection in the gain region and find that the PNS convection has very little impact on the general results we discuss here. 

\begin{figure}[tb]
  \centering
  \includegraphics[width = 0.47\textwidth]{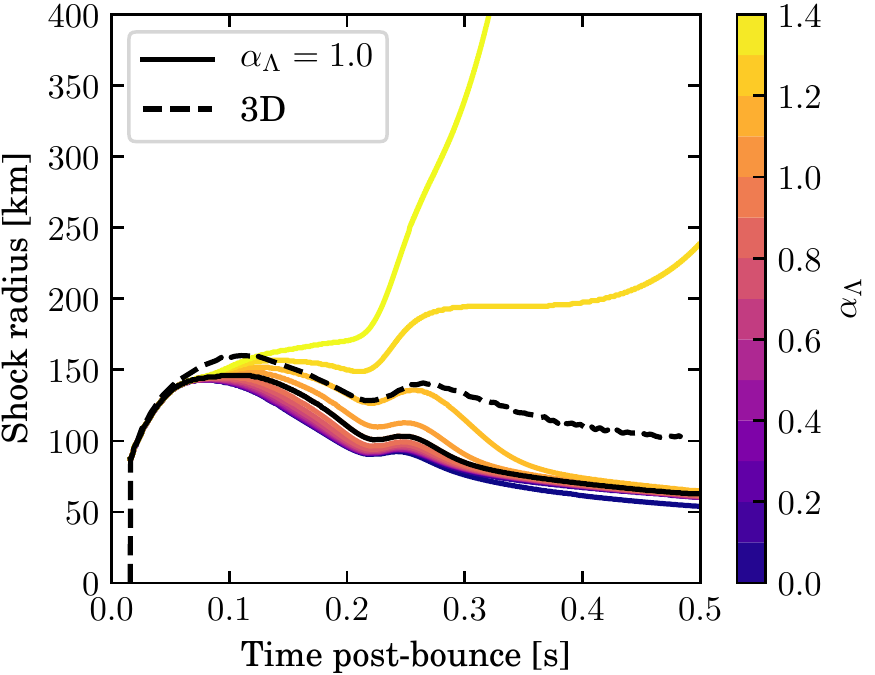}
  \caption{
    Shock radius evolution for our STIR simulations compared to the 3D simulation of \citet{oconnor:2018b}. 
    As the mixing length parameter \alphL is increased, the shock reaches larger and larger radii. 
    For $\alphL=1.3$ a successful explosion results in our 1D simulations.
    For $\alphL=1.2$ the shock radius closely follows the 3D evolution until $\sim$250 ms, at which time additional 3D effects such as SASI aid the shock in 3D and are not included in the 1D model.
    }
  \label{f.alpha_shock}
\end{figure}

\begin{figure}[htb]
  \centering
  \includegraphics[width=0.47\textwidth]{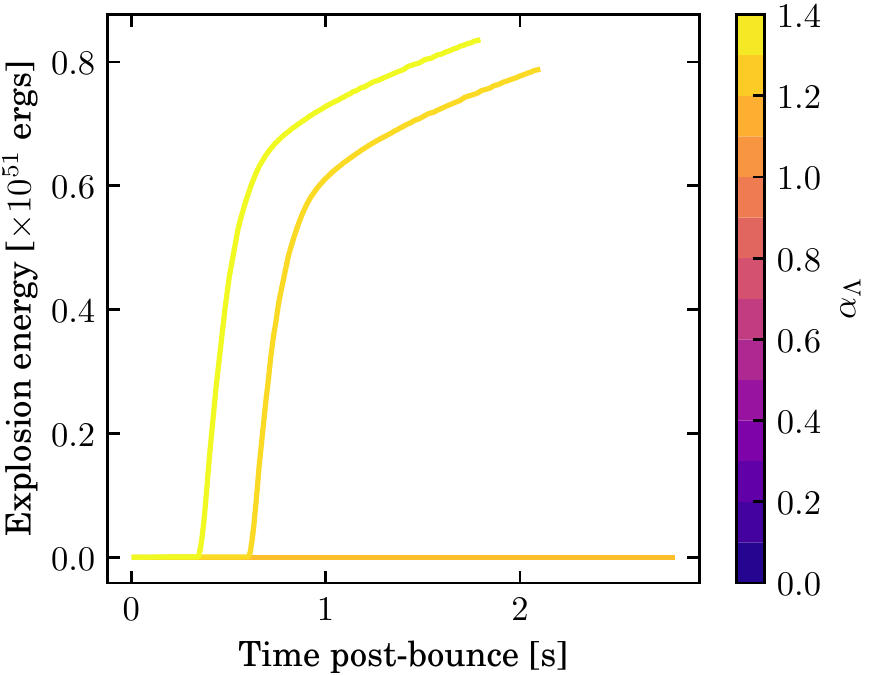}
  \caption{Diagnostic explosion energy as a function of time for 1D STIR simulations using the  20 \Msun progenitor model from \citet{oconnor:2018b}.
  This progenitor only explodes for \alphL values $\geq$1.3. }
  \label{f.alpha_ener}
\end{figure}

As evident from Figure \ref{f.alpha_vcon}, larger values of \alphL result in higher typical turbulent speeds and this has an important impact on the overall dynamics of the 1D CCSN simulation.
In Figure \ref{f.alpha_shock} we show the time evolution of the (average) shock radius from the 3D comparison simulation and from our STIR simulations with several values of \alphL.
At early times ($\lesssim 50$ ms), there is very little difference in the shock radii.
This is a result of the turbulent convection not being fully developed and strong at these early post-bounce times.
Once convection does become strong, which occurs slightly later in our STIR simulations than in the 3D simulation, the shock radius becomes a strong, increasing function of the \alphL. 
For sufficiently large values, a successful explosion results.
This is a consequence of the turbulent correlation terms included in Equations (\ref{e.mass3})-(\ref{e.ye}), most importantly the turbulent stress and dissipation terms.
For $\alphL=0$, all these additional terms are zero and the system of equations reduces to the usual 1D, spherically-symmetric case.
This is shown in Figure \ref{f.alpha_shock} as the darkest blue line with the smallest shock extension. 

In Figure \ref{f.alpha_ener} we show the diagnostic explosion energy \citep[see][]{bruenn:2013, muller:2012} as a function time post-bounce for our 1D STIR simulations using the 20-\Msun progenitor from \citet{oconnor:2018b}. 
These explosions energies are, evidently, still increasing slowly at the point we end our simulations. 
Here, our ability to capture the final, asymptotic explosion energies is limited by our assumption of NSE everywhere in the computational domain. 
This assumption becomes less and less correct at larger radii in the progenitors so we limit the radial extent of the domain to 15,000 km. 
We stop the simulations once the shock reaches this radial limit and the explosion energies are, in general, still increasing at this point, as seen in Figure \ref{f.alpha_ener}.
Therefore, we can only give an estimated lower limit for the diagnostic explosion energy of our STIR models. 
Beyond this, we see that there is a weak correlation with the rate of growth of the explosion energy and the \alphL value used. 
For failed explosions, which result for this progenitor for \alphL values of 1.2 and below, we run the simulations until PNS collapse, around 2.78 s for this model.

The \alphL value that most closely reproduces the shock evolution of the 3D simulation is around 1.2, which is also roughly the value that most closely reproduces the magnitude of the turbulent speeds from the 3D simulation (Figure \ref{f.alpha_vcon}).
At this \alphL the 1D STIR simulation results in a shock radius that tracks closely the 3D model of \citet{oconnor:2018b}, particularly prior to about 250 ms post-bounce. 
After this time, the STIR simulation shows a more rapid recession of the shock than the 3D simulation.
Other multidimensional effects that are not included in our STIR model, such as the SASI, may be aiding the 3D shock radius at these times.
The 3D simulation does not explode up to the 500 ms simulated by \citet{oconnor:2018b} and our STIR simulation with $\alphL=1.2$ also fails. 
We do not find an explosion for this progenitor with STIR until $\alphL=1.3$.
As we will see in Section \ref{s.param}, this is a very large ``critical'' \alphL for explosion.
As \citet{oconnor:2018b} found, it seems this progenitor is, indeed, stubbornly non-explosive.

\begin{figure}[tb]
  \includegraphics[width=0.47\textwidth]{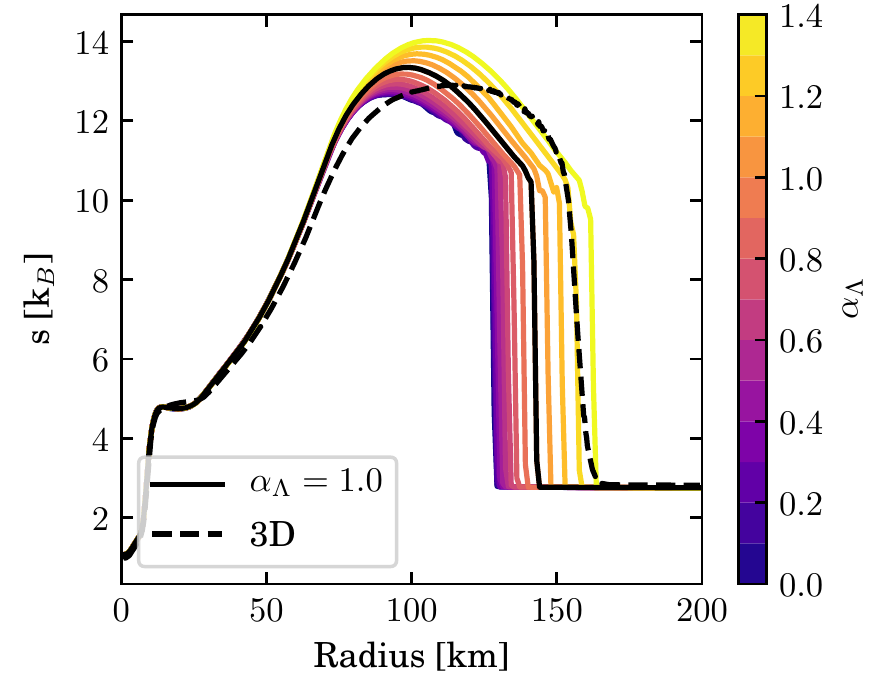} \\
  \includegraphics[width=0.47\textwidth]{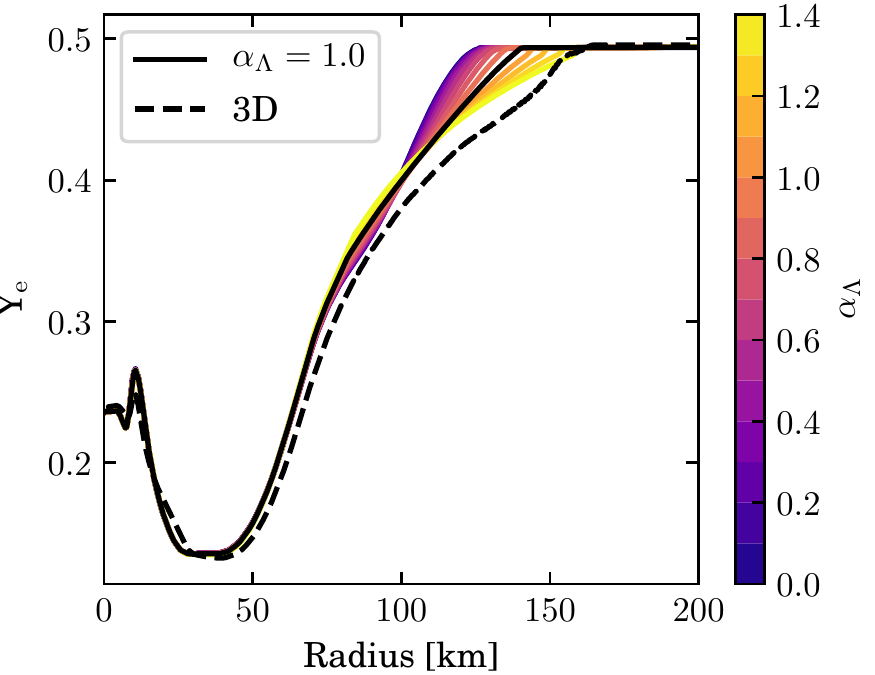}
  \caption{
    Entropy and \Ye radial profiles from our STIR models compared with angle-averaged radial profiles of the same progenitor from the 3D simulation of \citet{oconnor:2018b} at 135 ms post-bounce.  
  }
  \label{f.profiles}
\end{figure}

Figure \ref{f.profiles} shows the entropy and \Ye profiles from STIR for several values of \alphL compared to the corresponding angle-averaged profiles from the 3D simulation of \citet{oconnor:2018b} at 135 ms post-bounce.
Evident from the entropy profiles is the greater extension of the shock radius as \alphL is increased. 
The comparison of entropy profiles shows that, as implied by Figures \ref{f.alpha_vcon} and \ref{f.alpha_shock}, an \alphL value between 1.2 and 1.3 reproduces the average shock radius of the 3D simulation. 
Generally, the entropy profiles in the gain layer from STIR are steeper than those from the 3D simulation. 
Larger values of the diffusive mixing parameters in STIR can lead to flatter entropy profiles but overall we find a better agreement in the gross dynamics (i.e., shock radius, total turbulent energy) with the parameter values given above. 
The case is similar for the \Ye profiles. 
While increasing \alphL leads to profiles closer to the 3D case, the gain region \Ye values from STIR are always larger than for the comparable 3D simulation. 
Enhanced mass scalar diffusive mixing can improve this comparison. 
At small radii, around 10 km, the inability of STIR to accurately capture PNS convection is clear as the peak in \Ye is substantially higher than for the 3D case and the entropy profiles are slightly steeper. 
Adjusting the MLT diffusion parameters in STIR, particularly that for the compositional or trapped neutrino mixing (cf.  Equations (\ref{e.diffYe}) and (\ref{e.diffNu})) can improve the comparison to 3D, but more likely is that the use of the Ledoux criterion without the inclusion of the impact of trapped neutrinos is also hampering our ability to capture PNS convection at present \citep{roberts:2012a}.  
Exploration of this is left to future work.

\begin{figure*}[t]
  \centering
  \includegraphics[width=7in]{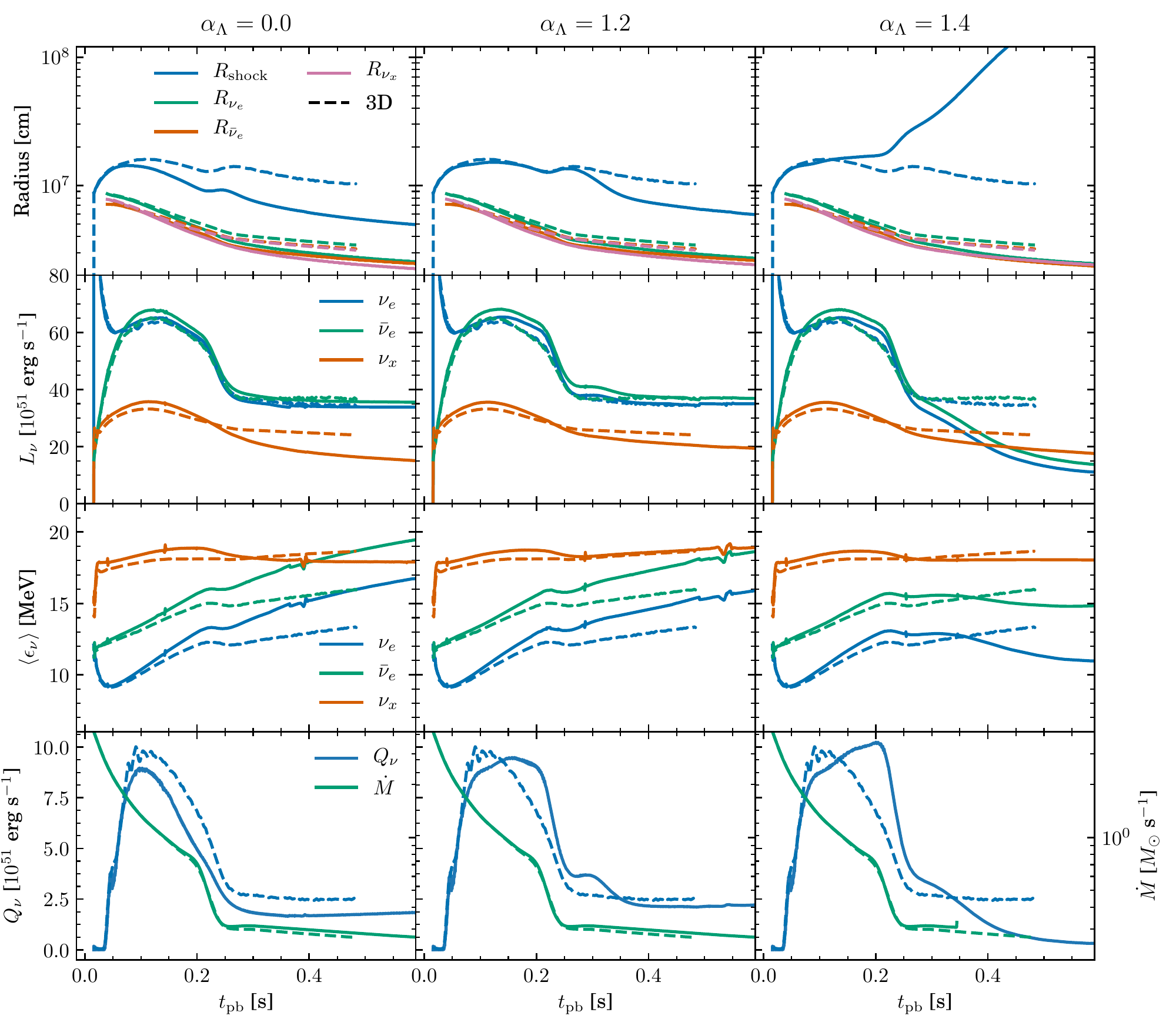}
  \caption{Comparison of key time-dependent metrics between STIR and the 3D simulation of \citet{oconnor:2018b}. STIR data are shown with solid lines while the 3D data are displayed with dashed lines. The left column shows data for \alphL= 0, the middle column shows \alphL = 1.2, and the right column shows \alphL = 1.4. The top row shows the shock radius, and the radii of the neutrinospheres for electron neutrinos, electron antineutrinos, and heavy lepton neutrinos. As \alphL is increased to 1.2, the shock radius evolution more closely matches the 3D simulation. For \alphL = 1.4, the 1D STIR simulation explodes successfully. The second row shows the neutrino luminosities. Increasing \alphL can slightly increase the luminosities of all three flavors we evolve until explosion occurs. The cessation of accretion accompanied by explosion dramatically reduces the electron neutrino and antineutrino luminosities. The third row shows the mean spectral energies of the neutrino emission. Increased \alphL hardens the spectra of heavy lepton neutrinos/antineutrinos while softening the spectra of electron-type neutrinos/antineutrinos since the convection causes the matter temperature to be higher at the heavy-lepton neutrinosphere. The fourth row shows the net heating rate in the gain region and the mass accretion rate at 500 km in radius. Similar to the neutrino luminosities, increasing \alphL enhances the neutrino heating, until explosion sets in. The mass accretion rates between 1D and 3D are essentially identical (until explosion occurs for \alphL = 1.4), with small differences arising just from the way that we compute an average accretion rate at 500 km in the Cartesian 3D grid used by \citet{oconnor:2018b}.}
  \label{f.mesa20det}
\end{figure*}

Figure \ref{f.mesa20det} shows a detailed comparison between our 1D STIR models and the 3D simulation for several key metrics and for three different values of the mixing length parameter \alphL. 
For \alphL = 0 (i.e., no inclusion of any turbulent convection in the 1D simulation), the shock radius after the 50 ms post-bounce remains substantially below that of the 3D simulation. 
The radii of the various neutrinospheres also decrease faster. 
The luminosities of electron neutrinos and antineutrinos are quite similar, though the heavy lepton neutrino luminosity is slightly higher at early times but falls off much faster than the 3D case. 
The mean energies of the neutrinos is fairly similar prior to about 150 ms, but then diverged significantly. 
After this time, in the 1D case, the mean energies of the electron neutrinos and antineutrinos increase rapidly while the mean energies of the heavy lepton neutrinos decrease. 
This is not seen in the 3D simulation where the energies of all neutrino flavors continue to increase, though the electron types increase in energy more slowly than for the \alphL = 0 1D case.
The net neutrino heating rate in the 1D \alphL = 0 case is less than that for the 3D simulation after about 75 ms post-bounce while the mass accretion rates are essentially identical.

The middle column of Figure \ref{f.mesa20det} compares the 3D simulation to the \alphL = 1.2 STIR simulation. 
This value of \alphL approximately fit the convective velocity, shock radius, and angle-averaged profiles of the 3D simulation fairly well. 
Prior to the accretion of the Si/O interface around 250 ms, the shock radius and neutrinosphere radii match the 3D simulation closely. 
The neutrino luminosities for the 1D STIR model are typically a bit higher than the 3D simulation. 
The neutrino mean energies are closer to those of the 3D simulation, particularly for the heavy lepton neutrinos, though the electron neutrinos/antineutrinos are slightly harder following the accretion of the Si/O interface.
The net heating rate for the \alphL = 1.2 case is more similar to that of the 3D case, though there is a slight deficit of heating prior to about 150 ms and a slight excess thereafter. 

The right column of Figure \ref{f.mesa20det} compares the \alphL = 1.4 STIR case to the 3D simulation. 
For this value of the mixing length parameter the 1D simulation successfully explodes. 
Up to around 100 ms, the shock radii between this 1D and the 3D simulation are very similar, then the 1D shock begins to expand while the 3D recedes. 
The shock expansion in the 1D case is accelerated when the Si/O interface is accreted around 250 ms. 
The neutrino luminosities in the STIR model with \alphL = 1.4 are generally enhanced prior to the onset of explosion, at which point they drop dramatically.
This is due to the cessation of accretion onto the PNS and attendant release of gravitational binding energy as neutrino radiation. 
The mean neutrino energies are comparable between the 1D STIR and 3D simulations, until the 1D begins to explode.
Then, the electron neutrinos and antineutrinos begin to soften as do, to a lesser extent, the heavy lepton neutrinos. 
The heating rate in the 1D STIR simulation is significantly enhance compared to the 3D after about 150 ms.
As explosion sets in, around the time the Si/O interface is accreted, the heating rate in the 1D case drops precipitously, falling below even the heating rate of the \alphL = 0 case at late times. 
This is a direct result of the drop in accretion luminosity and expansion of the post-shock matter, lowering the efficiency of neutrino capture in the gain region.
By the end of this STIR simulation (around 1.75 s when the shock reaches the edge of the domain), the heating rate is around $10^{50}$ erg s$^{-1}$, which is almost exactly the time-rate-of-change of the diagnostic explosion energy shown in Figure \ref{f.alpha_ener}.

Figure \ref{f.mesa20det} also shows that the radii of the neutrinospheres depends on \alphL. 
For no convection or turbulence (\alphL=0, left panel), the recession of the PNS is quite a bit more rapid than for the 3D case. 
For \alphL=1.2, the descent of the neutrinospheres is slowed, becoming more similar to the 3D case. 
Since STIR results in weak, essentially absent PNS convection, we do not attribute this to increased turbulent pressure support near the neutrinospheres but instead attribute the slower contraction of the PNS to the slower recession of the shock radius.
Even in spherical symmetry these two quantities are tightly connected \citep{janka:2001a}.
This coupling of the shock and PNS radii is only valid prior to the onset of explosion during the quasi-hydrostatic phase. 

Overall, the STIR model is able to reproduce fairly well many of the gross features of the 3D simulation such as the shock radius evolution, the neutrino luminosities and mean energies, and the gain region heating rate.
STIR does not yield comparable turbulent velocities in the PNS convection,.
We reiterate that, in STIR, we make no ad hoc modifications to the neutrino physics or transport and include full, multigroup, multidimensional transport of neutrinos identical to what is used in the comparison 3D simulation, except for the diffusive mixing of trapped neutrinos included in the 1D model. 

Comparing our STIR models to the 3D simulation of \citet{oconnor:2018b}, we find that the ``best fit'' value of \alphL is between 1.2 and 1.3.
This is similar to the MLT parameters that \citet{muller:2016a} find compare well to 3D simulations of convective O shell burning in a massive star.
There, they define the MLT ``$\alpha$'' parameters slightly differently than we do, but their $\alpha_1$ essentially corresponds to \alphL. 
They find that $\alpha_1 = 1$ describes the angle-averaged convective properties of the full 3D convective O shell which is quite similar to the \alphL values we find fit well to convection in a 3D CCSN simulation.
In the analysis of their 3D simulation, \citet{muller:2016a} also use diffusive mixing parameters of $\alpha_D = 1/6$, just as we do in STIR.

Our best fit value for \alphL as compared to the full 3D simulation of 1.2-1.3 is quite a bit smaller than the value often used for MLT in stellar evolution calculations of 2.0 \citep{paxton:2013, sukhbold:2014}.
This value of \alphL is far too large for our CCSN simulations and would result in very poor agreement with the 3D simulation and, as we shall see in the following section, very poor agreement with CCSN population metrics such as the explosion fraction. 
It must be noted, however, that the choice for the mixing length parameter in stellar evolution simulations is generally made on the basis of producing a good model for the sun, and convection in the post-shock region of a nascent CCSN is quite a bit different from that in the solar envelope.
That the preferred value is different by about a factor of two is not really that surprising but does, perhaps, serve to reiterate the concern that a single value for \alphL for all times in all places during stellar evolution is likely incorrect.
The constancy of \alphL, however, is a basic assumption of MLT which we also adopt here.
Over the few seconds of CCSN dynamics we simulate, during which time the fundamental nature of the convection does not change dramatically, this is probably not a terrible assumption.

\begin{figure*}[tb]
  \centering
  \includegraphics[width=\textwidth]{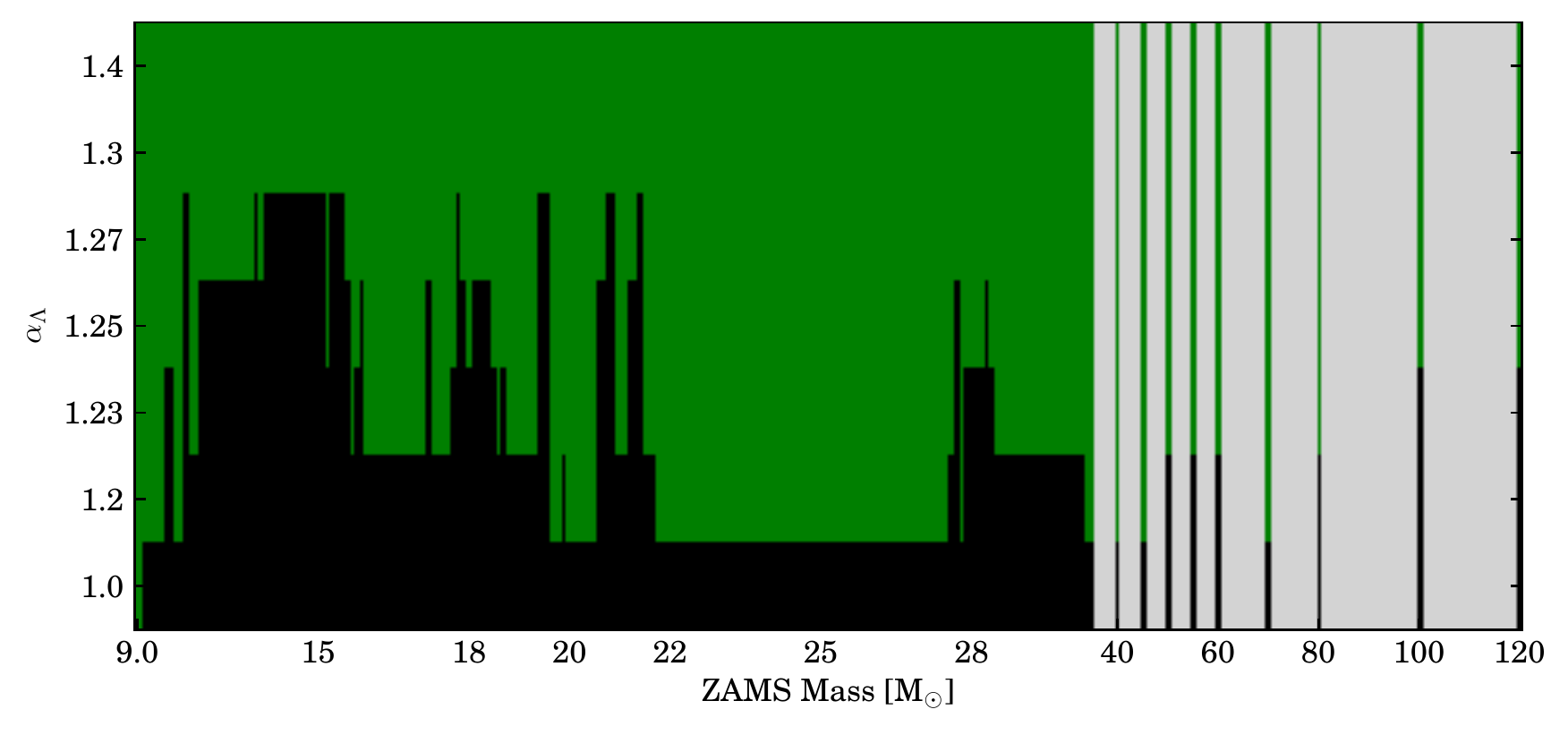} 
  \caption{Landscape of explosions for progenitors with ZAMS masses of 9-120 M$_{\odot}$ from \cite{sukhbold:2016} as a function of \alphL.  Green denotes successful explosion within the simulated time and black denotes failed explosion. Gray regions denotes the lack of a progenitor model.}
  \label{f.sukh_expl}
\end{figure*}

\begin{figure}[tb]
  \centering
  \includegraphics[width=0.47\textwidth]{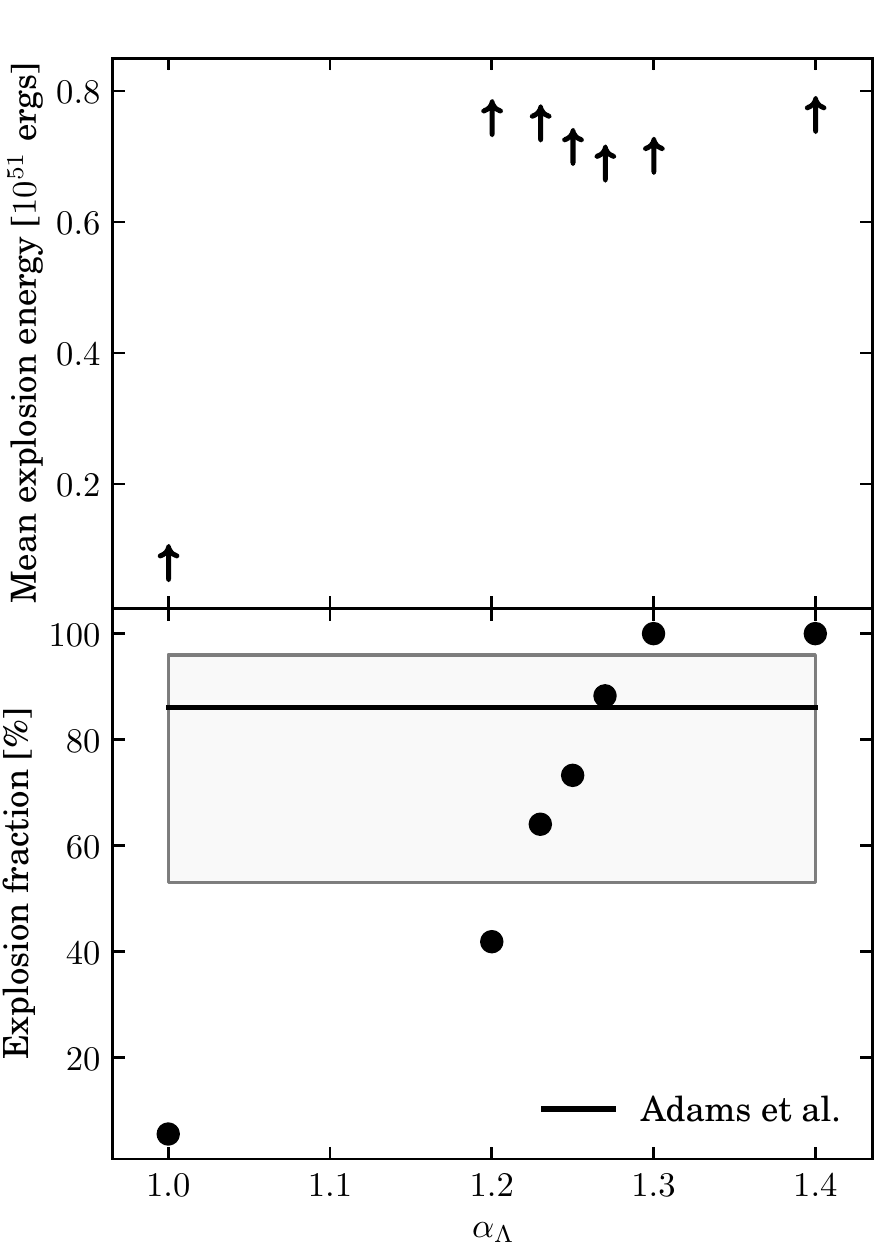}
  \caption{Average explosion energy of CCSNe for for progenitors from \cite{sukhbold:2016}, convolved with Salpeter IMF (top panel), and the total explosion fraction for our models also convolved with the IMF (bottom panel). The black line and error box in the bottom panel shows the observationally estimated explosion fraction of \citet{adams:2017}.}
  \label{f.sukhfrac}
\end{figure}

\section{CCSN Explosions from a Range of Masses} \label{s.param}

\subsection{Explodability and Explosion Energies}
\label{s.expl}

Having presented the formalism of STIR in Section \ref{s.turb} and shown that it can respectably reproduce the overall dynamics of a full 3D CCSN simulation in Section \ref{s.stir3d}, we now turn to an exploration of the explodability and observable characteristics for CCSNe arising from 200 progenitor models in the ZAMS mass range of 9-120 \Msun. 
Here we utilize the progenitor set of \citet{sukhbold:2016}, which is a superset of models from \citet{woosley:2007} and \citet{sukhbold:2014}.  We have used the whole range of progenitor masses (9-120M$_{\odot}$) to explore the sensitivity to progenitor mass.  
All of these progenitors are non-rotating, non-magnetic, solar metallicity, single stars.\footnote{Thus, no stars such as these likely exist in nature.}   
We refer the reader to \citet{sukhbold:2016}, and references therein, for detailed discussion of these progenitor models.
Our computational approach for these simulations is the same as that described above in Section \ref{s.num}.
For this first study, besides progenitor mass, we only vary the mixing length parameter \alphL of our STIR model, adopting several values from 1.0 to 1.4.
 
Figure \ref{f.sukh_expl} displays the ``explodability'' for the progenitor stars we study at several discrete values of \alphL. 
Here, by explodability we simply mean whether or not the star explodes (green in the figure) or fails (black).
We consider a model to have exploded if it has attained a net positive diagnostic explosion energy \citep[see, e.g.,][for definition]{muller:2012, bruenn:2016}.
For failed explosions, we run the simulations for up to 5 s post-bounce. 
For the vast majority of the failures, this is late enough to capture the onset of general relativistic instability of the PNS and collapse to a BH. 
For explosions, we run the simulations until the shock reaches the outer computational boundary.

A number of interesting features stand out in Figure \ref{f.sukh_expl}.
As has come to be expected \citep[cf.][]{muller:2019}, the low end of the mass range (9-10 \Msun) explodes more readily than more massive stars, with these progenitors exploding already for $\alphL=1.0$.
Beyond this low-mass window, the explodability of the progenitors as a function of ZAMS mass is extremely non-monotonic, with the jagged, sawtooth-like pattern described as ``islands of explodability'' by \citet{sukhbold:2016}.
Note, however, that while we find qualitative agreement about the complicated dependence of explodability on ZAMS mass between STIR and previous works \citep{ugliano:2012, sukhbold:2016, ebinger:2019}, detailed comparison shows that STIR predicts a somewhat different landscape of explosions.
We will return to a detailed discussion of this in Section \ref{s.criteria}.

\begin{figure*}[tb]
  \centering
  \includegraphics[width=\textwidth]{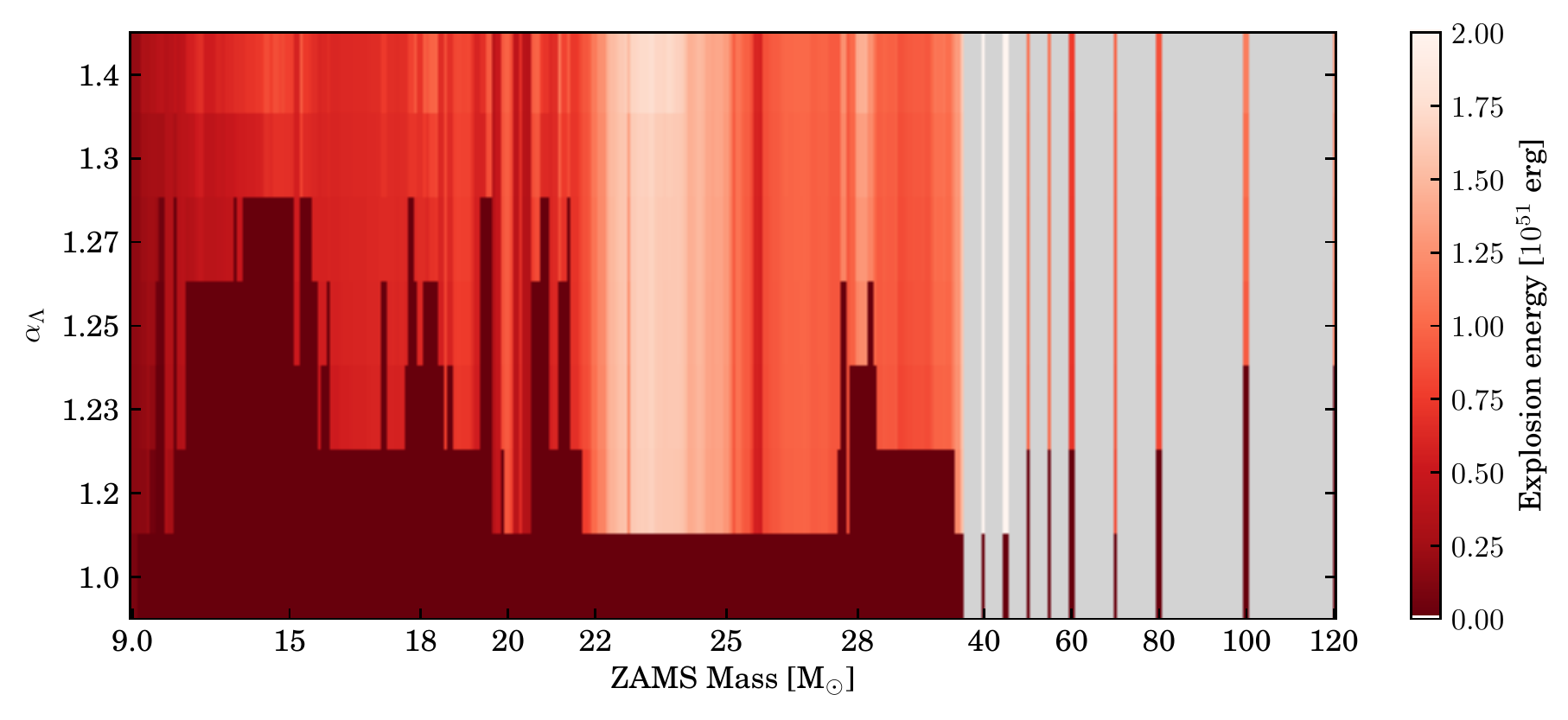}
  \caption{Explosion energies from STIR for progenitors with ZAMS masses of 9-120 M$_{\odot}$ and a range of \alphL = 0.0-1.4.  The color bar indicates the energy of the explosion, with black indicating that the simulation did not explode or the absence of a progenitor model at a given mass.}
  \label{f.sukh_expl2}
\end{figure*}

\begin{figure}[tb]
  \includegraphics[width=0.47\textwidth]{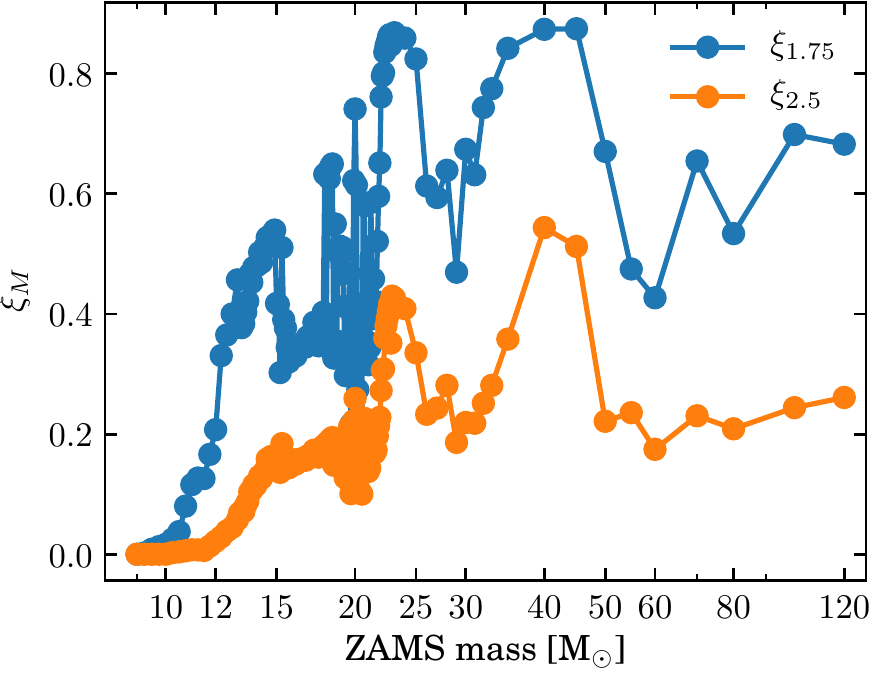}
  \caption{Core compactness at the onset of collapse as defined by Equation \ref{e.compact} for the progenitor model set of \citet{sukhbold:2016}. Shown are the 1.75 \Msun and 2.5 \Msun compactness values.}
  \label{f.compact}
\end{figure}

\begin{figure*}[htb]
  \centering
  \includegraphics[width=\textwidth]{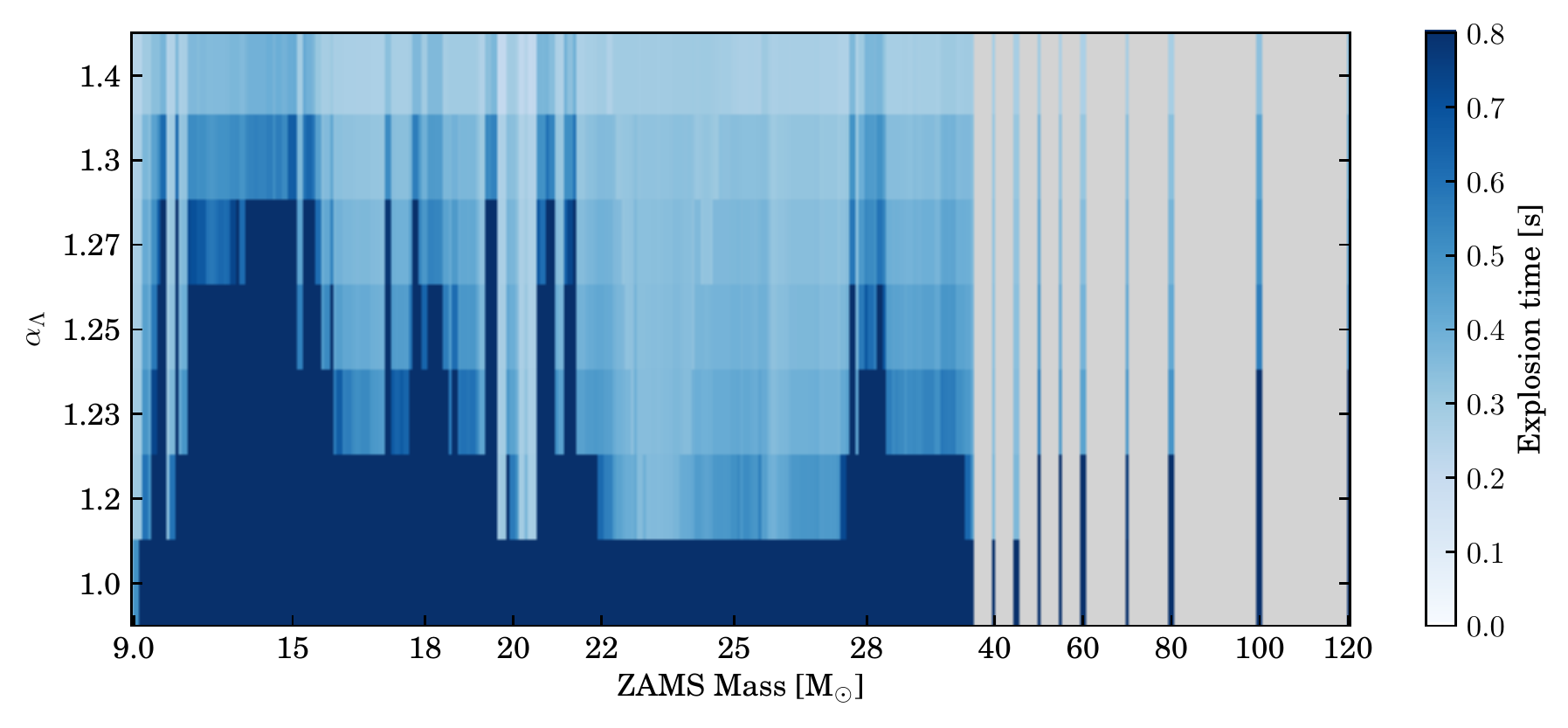}
  \caption{Explosion timescale of core-collapse supernovae with ZAMS masses of 9-120 M$_{\odot}$ and a range of \alphL = 1.0-1.4 for the \cite{sukhbold:2016} progenitor set.  The color bar indicates the timescale of the explosion, with black indicating the simulation did not explode. }
  \label{f.sukh_expltime}
\end{figure*}

\begin{figure}[tb]
  \centering
  \includegraphics[width=0.47\textwidth]{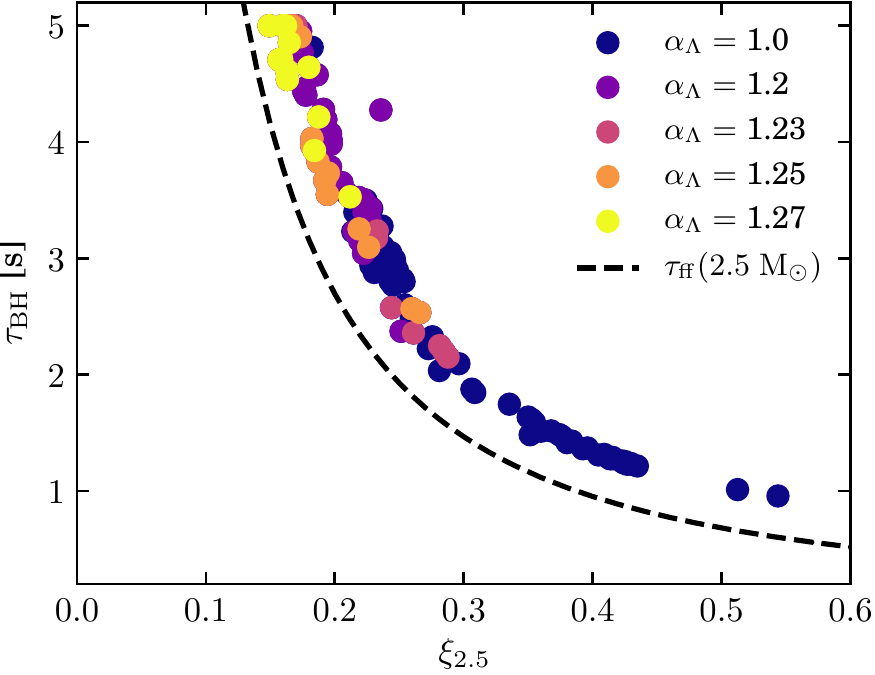}
  \caption{Black hole formation time versus compactness parameter $\xi_{2.5}$ for a range of \alphL values.  The dashed line shows the free fall timescale of the 2.5 \Msun mass shell \citep{oconnor:2011}.}
  \label{f.bhtime}
\end{figure}

The overall fraction of explosions from this set of progenitors is very sensitive to the mixing length parameter \alphL. 
In the bottom panel of Figure \ref{f.sukhfrac} we show the total explosion fraction as a function of \alphL weighted by a Salpeter initial mass function (IMF).
The explosion fraction is 6\% at $\alphL = 1.0$ and rises to $\sim$40\% at $\alphL = 1.2$
A linear increase in the explosion fraction is seen between $\alphL = 1.2$ and 1.3 to 100\%.
This sensitive behavior of explodability with \alphL is to be expected since, upon inspection of the evolution equations in Section \ref{s.turb}, we see that most of the turbulent correlation terms depend {\it at least} quadratically on \alphL.
This behavior also appeared for the comparison to 3D in Section \ref{s.stir3d}.
There, Figure \ref{f.alpha_shock} shows how the shock expansion depends non-linearly on \alphL and quickly transitions to explosion.
Also plotted in the lower panel of Figure \ref{f.sukhfrac} is the error box for the observationally determined explosion fraction from \citet{adams:2017},\footnote{Technically, \citet{adams:2017} estimate a ``failed'' supernova fraction. Here, we assume that the sum of the explosion fraction and the failed fraction is one.} along with their most likely value of $\sim$86\% (black horizontal line).
Values of \alphL that fall in the observational explosion fraction error box are 1.23, 1.25, and 1.27. 
These values of \alphL are also roughly those that compare best to the 3D simulation (Section \ref{s.stir3d}). 
We note, however, that our explosion fractions here do not include any stars with mass less than 9 \Msun that may evolve to core collapse. 

The top panel of Figure \ref{f.sukhfrac} shows the IMF-weighted explosion energy for the population of CCSNe produced by our progenitor set as a function of \alphL.
We show these data as lower-limits on the final predicted explosion energies because, in almost all cases, the explosion energy is still increasing by the end of our simulations (see Figure \ref{f.alpha_ener}).
We also do not include the ``overburden'' energy required to unbind the outer layers of the progenitor stars \citep[cf.][]{bruenn:2016}. 
The overburden can be quite significant, particularly for high compactness models. 
Due to this overburden, the final diagnostic explosion energy itself is an upper limit on the final explosion energy.
One drawback of STIR, like all 1D models, is that once an explosion is established, further accretion onto the PNS is shutoff, dramatically reducing the total emergent neutrino luminosities and attendant heating (see Figure \ref{f.mesa20det}). 
In 3D models, there is long-time continued aspherical accretion onto the PNS that helps maintain the neutrino luminosities \citep{lentz:2015, muller:2015a, muller:2019, vartanyan:2019}.
Unlike other 1D explosion parameterizations \citep[e.g.,][]{sukhbold:2016,perego:2015}, in STIR we do not add any sort of ad hoc late time enhancement of the neutrino luminosity or heating which might lead to a general inability of STIR to obtain as large explosion energies as comparable 3D simulations. 
Still, already at $\alphL = 1.2$ we find an average explosion energy approaching $8\times10^{50}$ erg, very near the canonical CCSN explosion energy of $\sim$10$^{51}$ erg.

Figure \ref{f.sukhfrac} shows that the IMF-weighted average CCSN explosion energy depends only weakly on \alphL. 
Of course, for $\alphL = 1.0$ where only low-mass progenitors explode, the average explosion energy is small. 
Note that in the averaging we only include models that actually explode. 
Low-mass progenitors systematically yield lower explosion energies in our models, in agreement with the 1D explosions in \citet{sukhbold:2016}. 
Once a large fraction of the progenitor models explode we see a constant population-average explosion energy with increasing \alphL.
For \alphL values of 1.3 and 1.4, for which all progenitors explode, there is a weak increase in average explosion going from 1.3 to 1.4. 
This indicates that {\it turbulence} plays a significant role in setting the explosion energy of our CCSN models. 
As pointed out by \citet{mabanta:2018}, and shown here by our STIR simulations, dissipation of turbulent kinetic energy to heat (cf. Equation (\ref{e.epsturb})) contributes significantly to the overall energy balance of the system (cf. Equation (\ref{e.ener3})). 
Larger values of \alphL naturally lead to larger amounts of turbulent kinetic energy behind the shock in what will become the CCSN ejecta. 
Once an explosion sets in, this turbulent kinetic energy is advected out with the ejecta (Equation (\ref{e.turbk})) and over time dissipates into thermal energy, contributing to the final explosion energy.
We find that after explosion occurs, while the base of the ejecta near the PNS is still subject to large neutrino heating, in the majority of the ejecta at later times the rate of dissipation of turbulent energy to heat far exceeds the rate of neutrino heating. 

Figure \ref{f.sukh_expl2} shows the landscape of explosion energies for our progenitor model set for each value of \alphL we run. 
Evident is the weak correlation with \alphL, but more apparent is the stronger non-monotonic dependence on progenitor ZAMS mass.
As discussed by previous works \citep[e.g.,][]{sukhbold:2016}, the explosion energy is most closely correlated with core {\it compactness}. 
The core compactness is defined as \citep{oconnor:2011},
\begin{equation}
  \xi_{M} =\left. \frac{M/\mathrm{M}_{\odot}}{R(M_{\mathrm{bary}} = M)/1000\mathrm{\, km}}\right|_{t = t_{\mathrm{collapse}}}, \label{e.compact}
\end{equation}
where $R(M_{\mathrm{bary}} = M)$ is the radius that encloses a mass $M$ in the progenitor star.
In \citet{oconnor:2011}, the authors compute the compactness at the point of core bounce.
Here, for convenience, we compute the compactness at the point of core collapse instead, which gives a very similar result \citep{sukhbold:2014}.
In Figure \ref{f.compact} we show the core compactnesses for $M$=1.75 \Msun and 2.5 \Msun for our progenitor model set \citep{sukhbold:2016}.
While we find that the vast majority of the model space results in somewhat weak explosions with energies $<10^{51}$, the highest compactness progenitors in the $23\ \Msun < M < 25\ \Msun$ range yield large explosion energies, upwards of $2\times10^{51}$ erg. 
These large energies are obtained even at the ``critical'' \alphL value for which these models just explode. 
The low-mass, low-compactness models that explode for \alphL values as small as 1.0 yield weak explosions. 
Even at $\alphL = 1.4$ these low-mass models barely reach explosion energies of $0.5\times10^{51}$ erg.  
We stress again, however, that these explosion energies are not the final, asymptotic energies that will be attained at later times than we are able to run these simulations and they do not include the overburden of the material above the shock. 
The variation in explosion energy with \alphL for a given progenitor is, in large part, due to the explosions occurring earlier at higher \alphL.

Figure \ref{f.sukh_expltime} shows the explosion times for our model set as a function of \alphL. 
We define the explosion time as the time when the diagnostic explosion energy exceeds 0.01$\times$10$^{51}$ erg. 
We find a wide range of explosion times from less than 0.1 s to more than 1 s, with explosions in most progenitors occurring before 0.5 s.
This is a range of explosion times very similar to what is found in multidimensional CCSN simulations (see \ref{s.comparison}) and significantly earlier than the explosion times for the 1D explosions parameterization of \citet{sukhbold:2016}, though similar to the explosion times from \citet{ebinger:2019}. 

The only ``time limit'' for explosion, physically, is the race against the collapse of the PNS to a BH. 
\footnote{Though, practically, very late explosions on average would significantly impact key observables such as nucleosynthetic yields in a manner inconsistent with observations \citep[cf.][]{woosley:1986b,woosley:1995a}.}
In STIR, since we include the full PNS with a realistic nuclear EOS and approximate general relativistic gravity, we are able to simulate the onset of PNS collapse to BH. 
Of course, once BH formation commences, our essentially Newtonian dynamics become wildly inappropriate and, furthermore, the central densities tend to exceed the limits of our EOS table. 
In Figure \ref{f.bhtime} we show our measured BH formation times for our model set as a function of compactness $\xi_{2.5}$. 
Our data for BH formation times match exceedingly well those from the fully general relativistic simulations of \citet{oconnor:2011}.
There, the authors suggest the free-fall timescale of the inner 2.5 \Msun of the progenitor as a good proxy for the BH formation time.
This time scale is also shown in Figure \ref{f.bhtime} and we also find it to be a reasonable approximation of our measure BH collapse times.

Figure \ref{f.bhtime} suggests that for progenitors more compact than about $\xi_{2.5} \sim 0.15$, 5 s is sufficiently long to capture the onset of collapse to BH, 
i.e., this is a long enough time scale to say definitively if a model will explode or not. 
For less compact progenitors, the BH formation time as approximated by the free-fall time of the 2.5 \Msun mass shell becomes considerably longer.
It is possible that these progenitors may explode at times later than we consider here. 
Still, with this caveat, we consider any model that has failed to explode within the $\sim$5 s we simulate to be a ``failed'' explosion that will result in BH formation. 
This time scale is substantially longer than any explosion time we find in our current study.

\subsection{Explosion criteria}
\label{s.criteria}

Understanding the impact of the population of CCSNe on, e.g., cosmic chemical evolution or to compare directly to observational data on things such as the compact object mass distributions, we need theoretical predictions from the wide range of potential progenitors of CCSNe. 
Given the enormous expense and complexity of multidimensional CCSN simulations, even in 2D, this is not yet feasible.
Thus, an area of perennial interest is to develop criteria for predicting which progenitor stars will explode, and where their explosion and remnant properties will be, based solely on the pre-collapse progenitor structure itself. 
In the present work, we refrain from developing a new analytic, or semi-analytic, explosion criterion based on STIR and instead restrict ourselves to a comparison of our STIR results to a few select explosion criteria. 

The first and easiest metric by which to attempt to predict explosion or failure is the ZAMS mass. 
ZAMS mass is still, regrettably, the introductory ``textbook'' differentiator between success and NS production or failure and collapse to a BH, with the typical cutoff somewhere around 25 \Msun \citep[cf.][]{heger:2003}.
A multitude of theoretical work in 1D, 2D, and 3D, however, has already shown this to be a poor criterion, finding explosions for stars more massive than 25 \Msun and/or failures for stars less than this \citep[e.g.,][]{oconnor:2011, ugliano:2012, sukhbold:2016, ertl:2016, summa:2016, roberts:2016, vartanyan:2018, vartanyan:2019, ott:2018, oconnor:2018, oconnor:2018b, ebinger:2019}.
As in these works, our results with STIR clearly show that the ZAMS mass is a poor indicator for which stars will explode, as gleaned from Figures \ref{f.sukh_expl} and \ref{f.sukh_expl2}. 
These figures show that for \alphL = 1.25, most of the high mass stars in our model set successfully explode. 
Even the oft-collapsed-to-BH 40 \Msun star explodes quite energetically.
This model has the highest compactness in the \citet{sukhbold:2016} set and, hence, a very short BH formation timescale \citep[see Section \ref{s.remnants} below and][]{oconnor:2011}, making it ideal for multidimensional studies of BH formation \citep{pan:2018, chan:2018}.
Yet, our STIR simulations show this model is not particularly non-explosive.
The 40 \Msun progenitor has a ``critical'' \alphL value of 1.2, making it one of the easiest models to explode in our current set. 
And even at this value of \alphL, it explodes quite robustly with the highest explosion energy we obtain (2.7$\times$10$^{51}$ erg).\footnote{Though the overburden energy of this progenitor is about 1.6$\times$10$^{51}$ erg, bringing the corrected explosion energy down much closer to the canonical 10$^{51}$ erg.} 
We note this is also a distinguishing feature between our STIR simulations and the 1D neutrino-driven explosion model of \citet{sukhbold:2016} and \citet{ertl:2016}.
There, the authors find the 40 \Msun model fails to explode, even with the most robust neutrino driving engine parameterization.
This model also explodes in the 2D simulations of \citet{pan:2018}, depending on the EOS used, as well as in the 3D general relativistic simulations of \citet{ott:2018}.
\citet{chan:2018} find a successful 3D explosion with this 40 \Msun progenitor accompanied by ``fallback'' BH formation.
Similar coincidental explosion and collapse of the PNS to a BH was observed in the 2D simulations of \citet{pan:2018} for the DD2 and LS200 equations of state. 
In our STIR models, the 40 \Msun model explodes without collapse to BH, though the final PNS mass is extremely close to the maximum mass limit for the SFHo EOS. 
Since we have to stop the simulation once the shock reaches the outer computational boundary (about 1.8 s post-bounce for \alphL = 1.25), we cannot say for certain that this model will not ultimately collapse to a BH at later times. 
Such cases of explosion with BH formation were seen in the 1D simulations of \citet{sukhbold:2016} when fallback accretion onto the PNS was accounted for.

Initial stellar mass is a poor distinguishing criterion between explosion and failure. 
Indeed, of stars with ZAMS mass greater than 25 \Msun only the 27.7- and 28.3-\Msun stars fail to explode at \alphL = 1.25 while the 120 \Msun progenitor explodes readily at \alphL = 1.25.
This non-monotonicity of explosion w.r.t. ZAMS mass will have important implications for, e.g., the distribution of compact remnant masses.
For instance, the majority of the BHs we find at \alphL = 1.25 come from the range of 13 to 16 \Msun, when accounting for the IMF.
These issues are discussed in greater detail below in  Section \ref{s.remnants}.

\begin{figure}[tb]
    \includegraphics[width=0.47\textwidth]{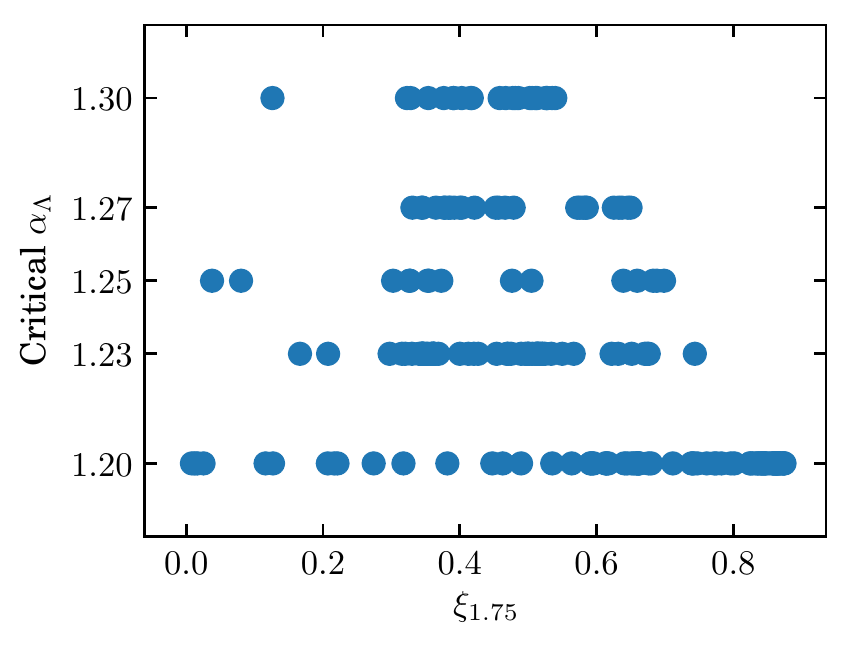}
    \caption{Critical \alphL, the lowest value of \alphL at which the progenitor explodes, versus progenitor core compactness $\xi_{1.75}$.}
    \label{f.critalpha}
  \end{figure}

\begin{figure*}[htb]
    \centering
    \includegraphics[width=\textwidth]{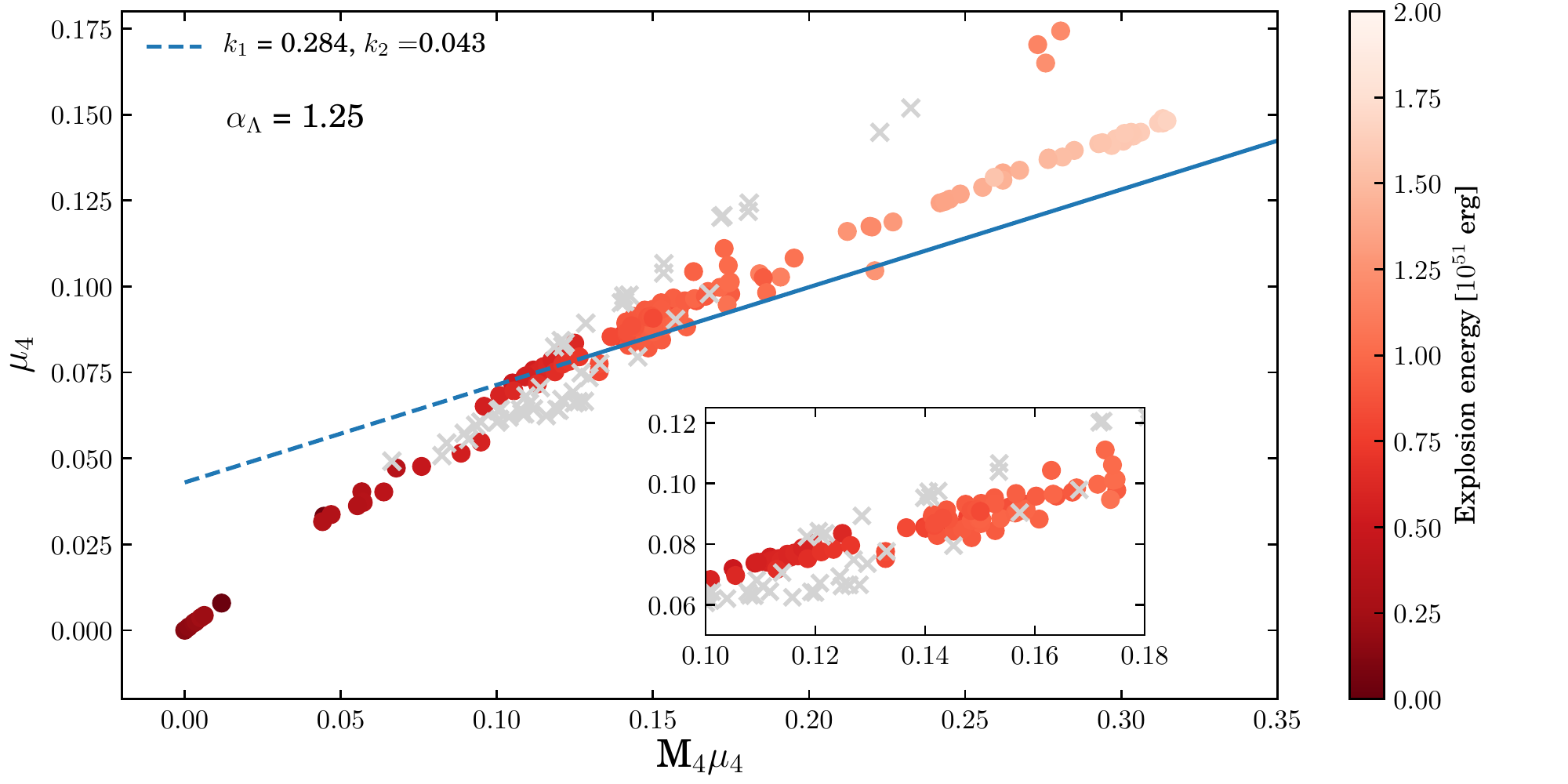}
    \caption{Two parameter explosion criterion from \citet{ertl:2016}. $\mu_4$ is the slope of the mass profile at the mass coordinate $M_4$ at which the entropy profile in the progenitor exceeds an entropy of 4 $k_B$ baryon$^{-1}$, as defined by Equations \ref{e.m4}) and (\ref{e.mu4}). Failed explosions are marked by gray crosses whereas successful explosions are circles, color-coded by their respective explosion energies. }
    \label{f.ertl}
  \end{figure*}

A major contributing factor to why ZAMS mass is such a poor criterion for predicting the outcome of stellar core collapse is that, due to the complex processes of stellar evolution, it is largely disconnected from the final structure of a massive star's core at the point of collapse.
Even in the case of single stars at solar metallicity, the highly non-linear physics of radiation hydrodynamics coupled to nuclear burning, which becomes more and more sensitive to small variations in the thermal structure of the star as evolution proceeds, combine to make the outcomes of stellar evolution almost {\it chaotic} as a function of ZAMS mass \citep{sukhbold:2014}. 
\citet{oconnor:2011} pointed out that a far better indicator for many aspects of stellar core collapse is the compactness of the final (usually iron) core (see Equation (\ref{e.compact})).

In Figure \ref{f.critalpha} we show the ``critical'' value of \alphL at which a progenitor model just explodes plotted against the compactness for a mass of 1.75 \Msun. 
\citet{oconnor:2013} argue this value is more closely related to the early post-bounce behavior since it is the typical baryonic mass enclosed by the shock soon (200-300\,ms) after bounce.
Our STIR models show there is a weak linear relationship between the critical \alphL value (i.e., the explodability) and the compactness $\xi_{1.75}$ above values of about 0.6 in the {\it opposite} direction: for high-compactness progenitors, higher compactness corresponds with lower critical \alphL values.
The very lowest compactness progenitors also explode readily, some already at $\alphL=1.0$, but besides these two extreme ends of the compactness scale there is no correlation at all with compactness and explodability.
In order to draw more salient conclusions about the relationship between compactness and explodability, we would clearly need to sample \alphL more finely in our parameter study. 
Our present data simply cover too few values of \alphL to say much conclusive beyond the correlation at high compactness.

\citet{ugliano:2012} and \citet{sukhbold:2016} also point out that the compactness on its own, while somewhat better than ZAMS mass, is a generally poor predictor of the success or failure of a given model. 
In the quest for a better still explosion indicator based solely on progenitor structure, \citet{ertl:2016} present a two-parameter criterion.
This criterion is based on the mass enclosed at the point where the entropy in the progenitor exceeds a value of 4 $k_B$ baryon$^{-1}$ and the value of the slope of the mass profile at this point.
Specifically, 
\begin{equation}
  M_4 \equiv m(s=4) / \Msun, \label{e.m4}
\end{equation}
and 
\begin{equation}
  \mu_4 \equiv \left. \frac{dm/\Msun}{dr/1000\ \mathrm{km}} \right\vert_{s=4}, \label{e.mu4}
\end{equation}
where $m$ is the enclosed mass, $r$ is the radius from the center of the star, and $s$ is the entropy in units of $k_B$ baryon$^{-1}$.
\citet{ertl:2016} argue that in the plane formed by $\mu_4$ vs. $M_4 \mu_4$ there is a distinct dividing line above which a progenitor model will fail and collapse to a BH and below which a successful neutrino-driven explosion will proceed. 
Based on parameterized 1D neutrino-driven explosion models along the lines of those in \citet{ugliano:2012} and \citet{sukhbold:2016} using 621 different stellar progenitor models, \citet{ertl:2016} find that their two parameter criterion predicts explodability with about a 97\% accuracy. 

In Figure \ref{f.ertl} we plot the results of our STIR simulations for \alphL = 1.25 in the $\mu_4$ vs. $M_4 \mu_4$ plane along with the steepest of the dividing lines between explosion (below) and failure (above) from \citet{ertl:2016}.\footnote{As in \citet{ertl:2016}, we do not use Equations (\ref{e.m4}) and (\ref{e.mu4}) directly but instead following the prescription described in their Section 3.2.} 
In this plot, failed explosions are denoted by gray crosses and explosions by colored circles where the color indicates the measured explosion energy of the model.
As can be seen, our results with STIR do not follow the same trend and separation between explosion and failure as those of \citet{ertl:2016}. 
We find many successful explosion above the separation line and many failures below it. 
Above $M_4 \mu_4 \sim 0.15$, our STIR results do seem to separate into two branches in the parameter space of Figure \ref{f.ertl}, the higher branch leading to failure and the lower branch leading to explosion.
This might seem to indicate that simply making the separation line proposed by \citet{ertl:2016} steeper might lead to a similarly good fit as those authors find.
This, however, does not hold at lower values of $M_4 \mu_4$. 
The inset in Figure \ref{f.ertl} shows a zoom-in of the low-$M_4 \mu_4$ region.
While a steep enough separation line might nicely divide explosions and failures at higher $M_4 \mu_4$, at smaller values it is the models with {\it higher} $\mu_4$ values at a given $M_4 \mu_4$ that explode.
Thus, no single curve in this plane would describe the outcomes of our STIR simulations to the degree of accuracy found by \citet{ertl:2016}.

Including the impact of turbulent convection in our 1D models clearly results in dynamics that are significantly different from those for purely neutrino-driven 1D explosions. 
Our STIR model most likely captures more closely the integral condition for explosion including the effects of turbulence by \citet{murphy:2017} and \citet{mabanta:2018}.
It is hard to compare our results directly to these works since there the authors have assumed a steady-state with constant neutrino luminosities and mass accretion rates onto the stalled shock.
\citet{murphy:2017} and \citet{mabanta:2018} approach the problem from the perspective of a ``critical neutrino luminosity'' as introduced by \citet{burrows:1993}.
Our approach is far more general, accounting for the detailed physics of the neutrino transport and time-dependent dynamics of core collapse for realistic progenitors making comparison between STIR and these other models challenging.
This is also the case for other explosion criteria such as the ``antesonic'' condition of \citet{pejcha:2012} and \citet{raives:2018}, which does not account for turbulence or convection.

\begin{figure}[tb]
  \centering
  \includegraphics[width=0.47\textwidth]{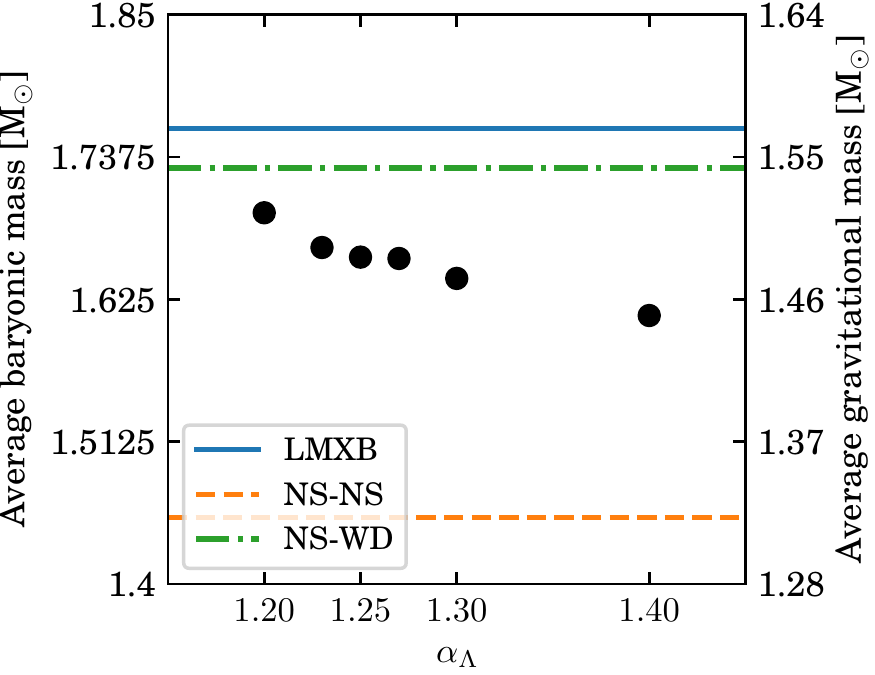}
  \caption{Average neutron star mass versus \alphL for the \cite{sukhbold:2016} progenitor set, convolved with the Salpeter IMF.  Horizontal lines indicate average neutron star masses as measured for low mass x-ray binaries (LMXB), double NS binaries, and NS-white dwarf binaries \citep[][stellarcollapse.org, 2 Februrary 2019]{lattimer:2012}. }
  \label{f.avgnsmass}
\end{figure}

\begin{figure*}[tb]
  \begin{tabular}{lll}
    \hspace{-0.25in}\includegraphics[width=0.33\textwidth]{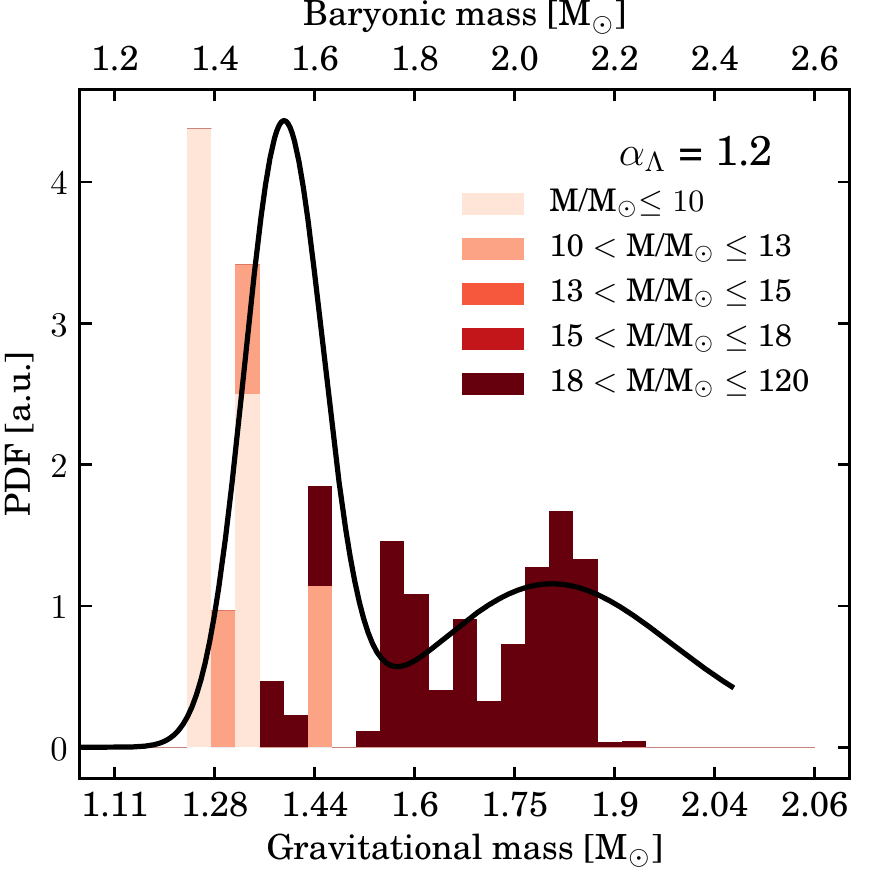} &
    \hspace{-0.2in}\includegraphics[width=0.33\textwidth]{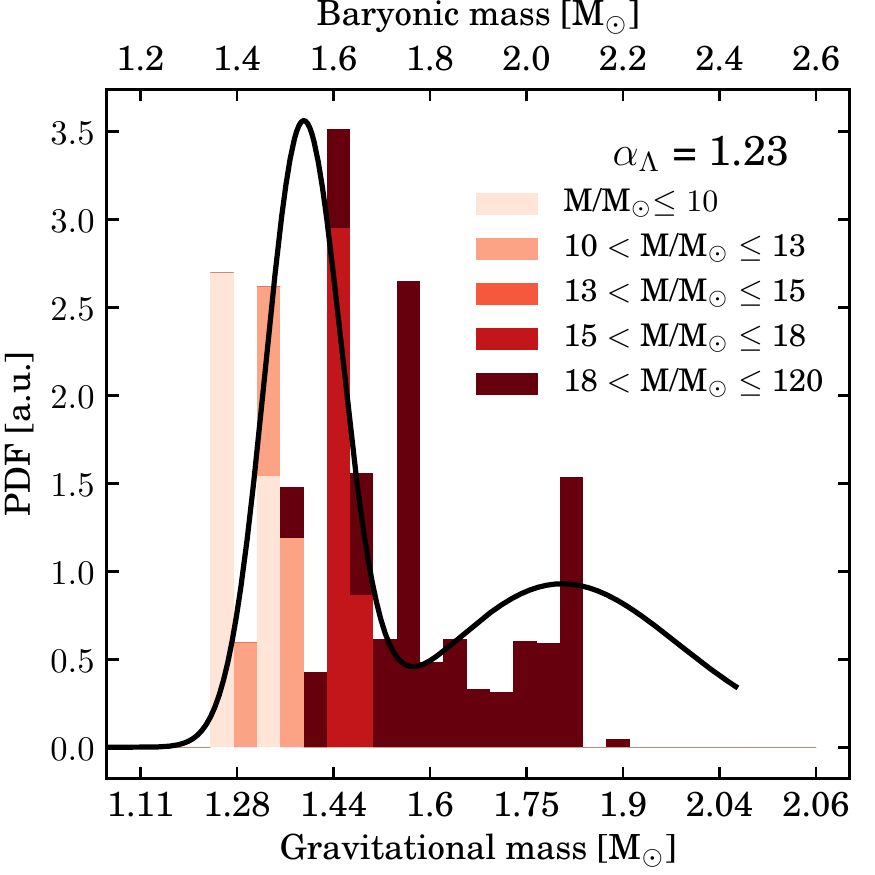} &
    \hspace{-0.2in}\includegraphics[width=0.33\textwidth]{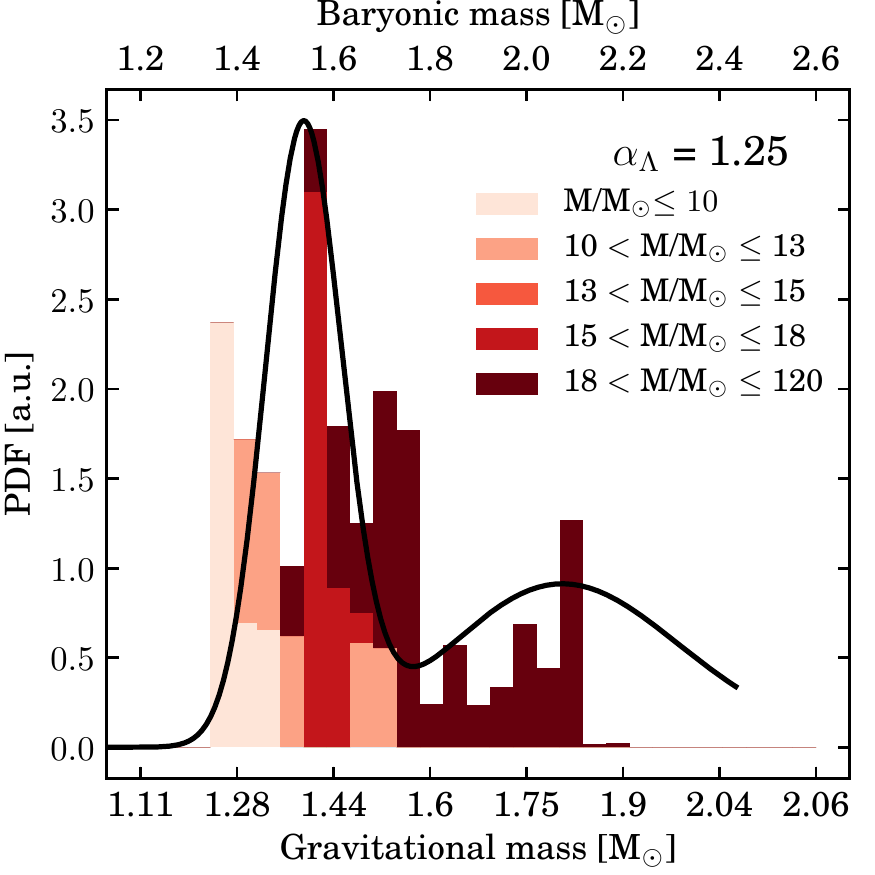} \\
    \hspace{-0.25in}\includegraphics[width=0.33\textwidth]{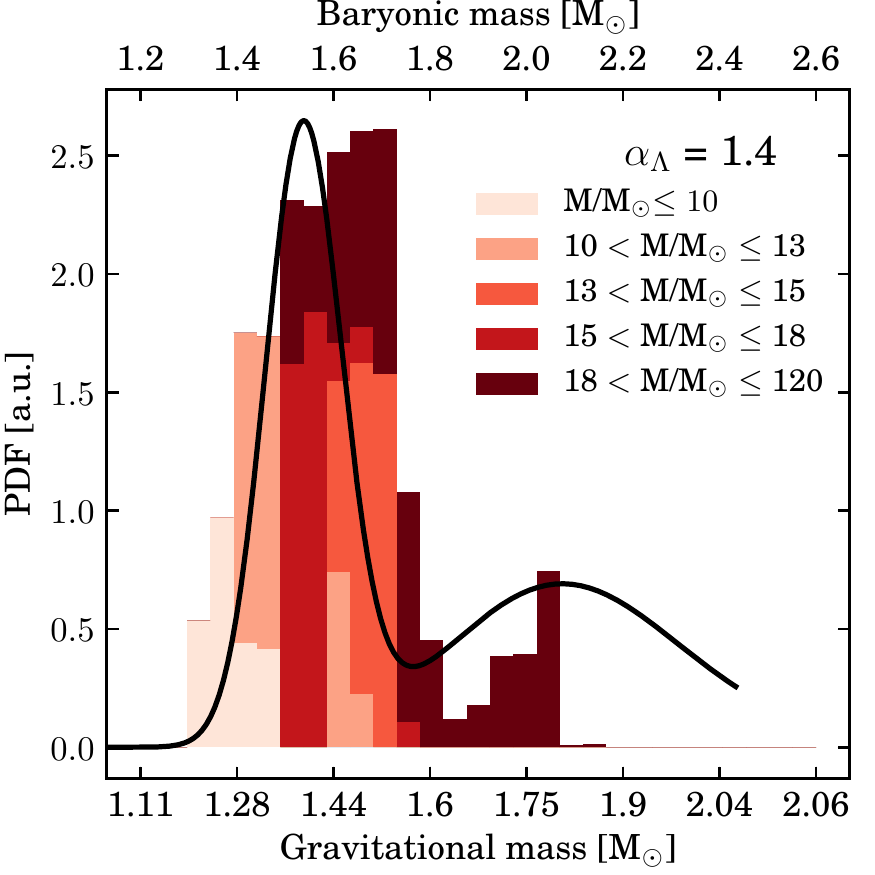} &
    \hspace{-0.2in}\includegraphics[width=0.33\textwidth]{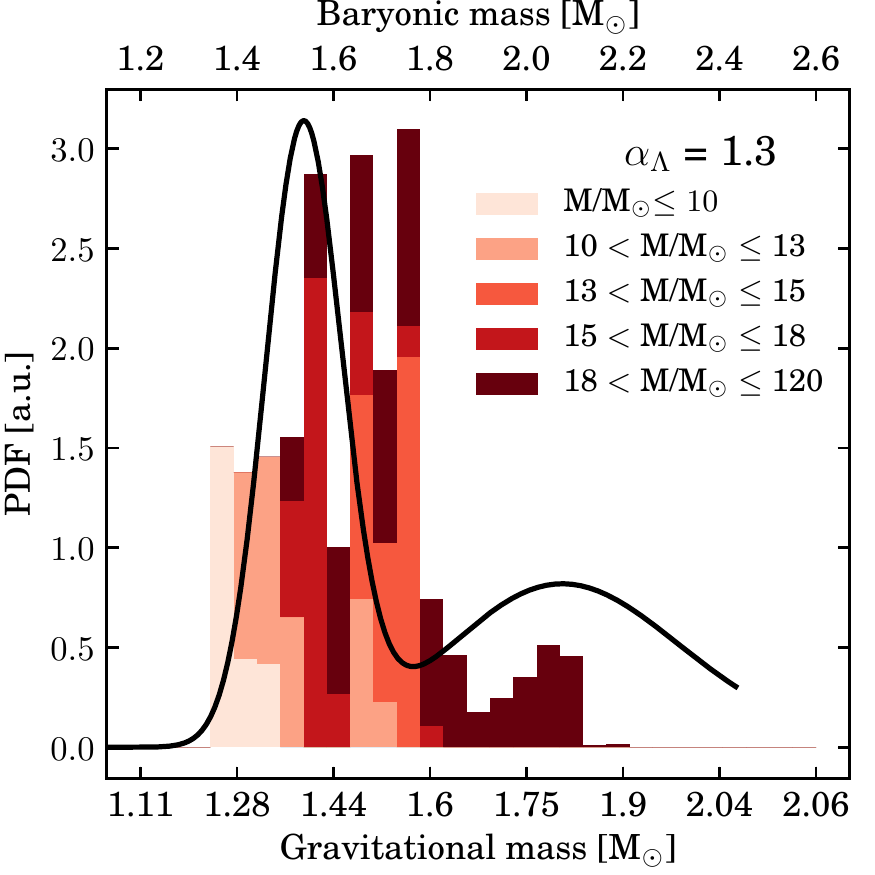} &
    \hspace{-0.2in}\includegraphics[width=0.33\textwidth]{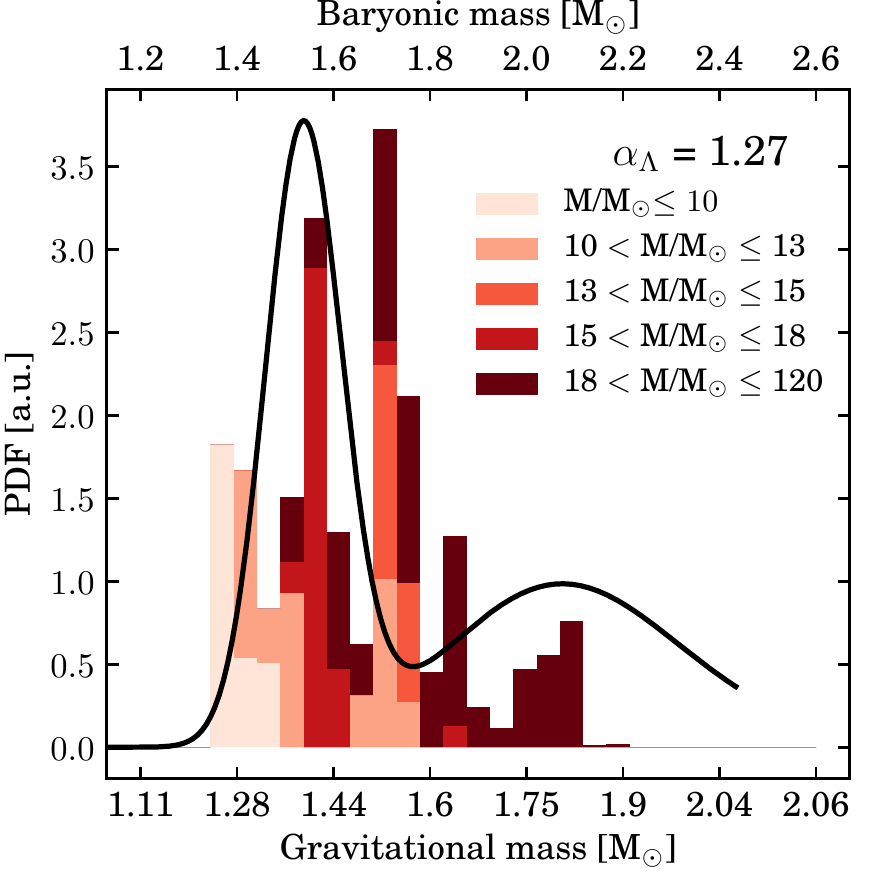}
  \end{tabular}
  \caption{Neutron star mass distribution for the \cite{sukhbold:2016} progenitor set, convolved with Salpeter IMF, for several values of \alphL.  Only simulations that produced successful explosions are included here. Top left is for \alphL = 1.2, top middle \alphL = 1.23, top right is \alphL= 1.25, bottom right is \alphL = 1.27, bottom middle is \alphL = 1.3, and bottom left is \alphL = 1.4.  Colors indicate ZAMS mass range of progenitor star.  The black line is the double-peaked model based on observed millisecond pulsar masses \citep{antoniadis:2016}. The values on the y-axis are the probabilities in arbitrary units (a.u.). }
  \label{f.nsdist}
\end{figure*}

\begin{figure*}[htb]  
  \begin{tabular}{lll}
    \hspace{-0.25in}\includegraphics[width=0.33\textwidth]{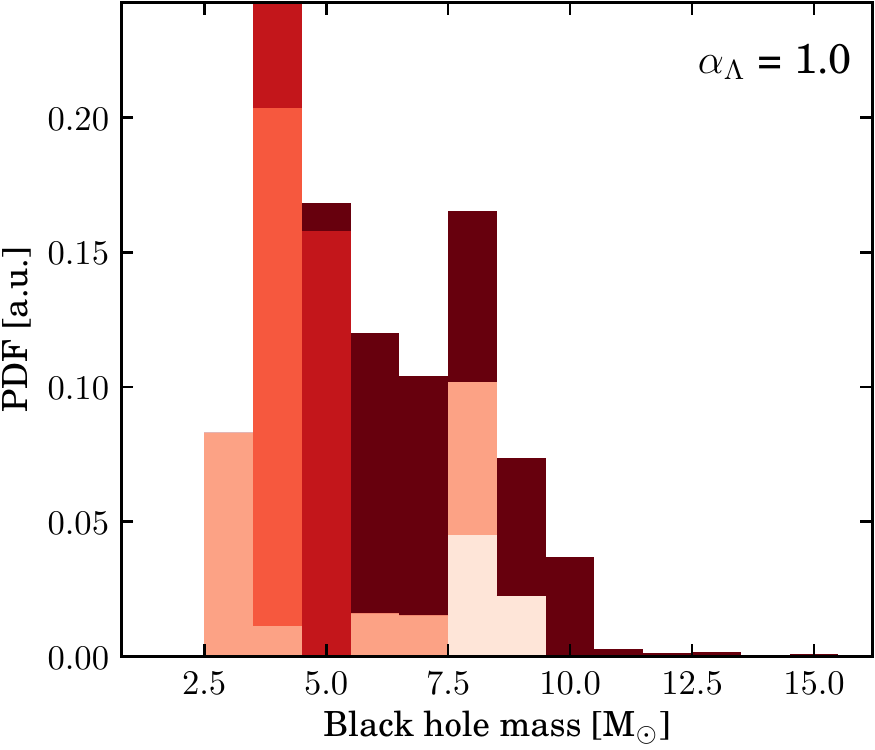} &
    \hspace{-0.1in}\includegraphics[width=0.33\textwidth]{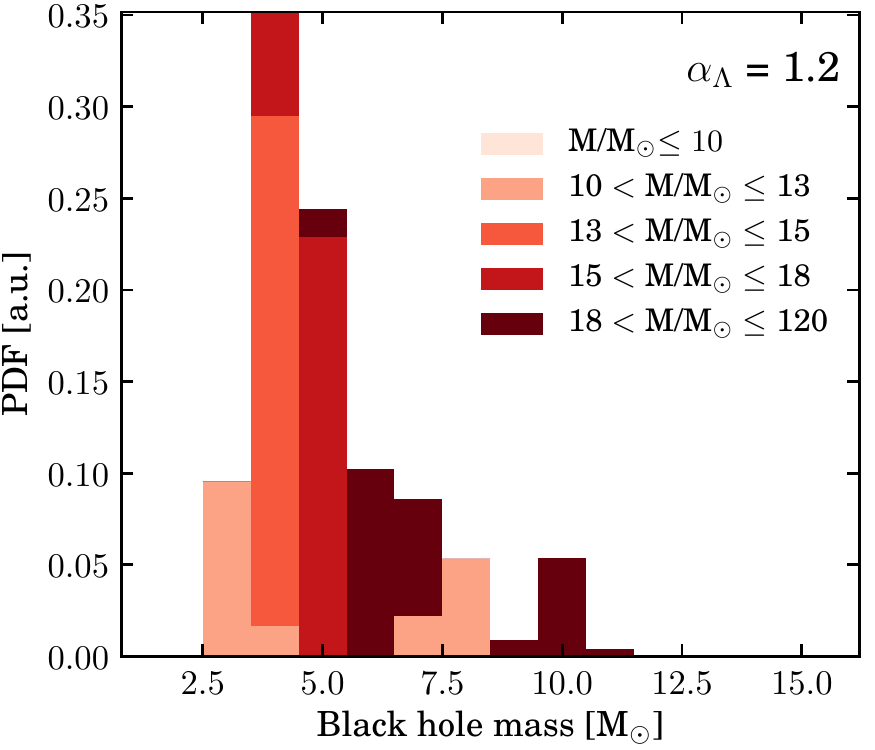} &
    \hspace{-0.1in}\includegraphics[width=0.33\textwidth]{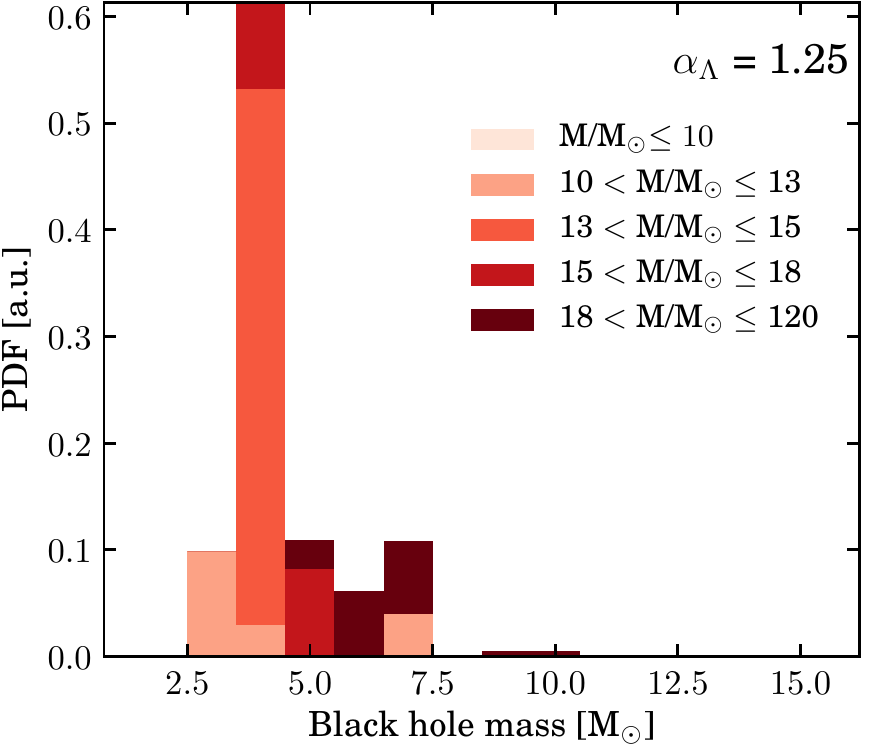}
  \end{tabular}
  \caption{Black hole baryonic mass distribution for the \cite{sukhbold:2016} progenitor set, convolved with the Salpeter IMF.  Only simulations that failed to explode are included here.  The left panel is for \alphL = 1.0, the middle for \alphL = 1.2, and the right for \alphL = 1.25. }
  \label{f.bhdist}
\end{figure*}

\subsection{Compact remnant masses}
\label{s.remnants}

We now turn to the masses of the compact remnants left behind by our STIR simulations. 
While the explosion energies in the vast majority of our successful models are still increasing by the end of the simulations (see Figure \ref{f.alpha_ener}), the PNS mass asymptotes quite quickly following the onset of explosive shock expansion due to the rapid cessation of accretion onto the PNS. 
In reporting the NS masses in what follows, we do not include any potential fallback that might add to the PNS mass at later times. 

Figure \ref{f.avgnsmass} shows the IMF-weighted average NS mass as a function of mixing length parameter \alphL. 
Shown are both the baryonic and gravitational PNS mass, the latter we compute by solving the Tolman-Oppenheimer-Volkoff equations for the SFHo EOS. 
Shown also on this Figure as horizontal lines are the observationally-derived average NS masses from three classes of objects \citep{lattimer:2012}: low-mass X-ray binaries (LMXBs), NS-NS binaries, and NS-white dwarf (WD) binaries.
Note that the width of the NS gravitational mass distributions for these different classes of systems are rather large: $\sim$1.0 \Msun for LMXBs, $\sim$0.8 \Msun for NS-WDs, and $\sim$0.3 \Msun for NS-NSs \citep[][Figure 8]{lattimer:2012}. 
We find a generally declining trend in average NS mass with \alphL.
This is the result of two factors.
First, the highest compactness progenitors explode with lower critical \alphL values (cf. Figure \ref{f.critalpha}) and result in the most massive NS's.
Second, the explosions occur earlier for higher \alphL leading to less mass NS's since accretion on the PNS is shutoff earlier. 
The NS masses we find at all of the shown values of \alphL are right in the range of observationally-determined average PNS masses from \citet{lattimer:2012}, especially when consider the large spreads in the observed distributions.
Compared to the best constrained average derived from double NS systems of $\sim$$1.3 \pm 0.15$ \Msun, the average NS masses we find are perhaps a little high, though we consider only single star progenitors and binarity could substantially impact the population-averaged NS masses.

As with all integration, the population-averaged NS mass contains less information than the underlying distribution of NS masses.
The detailed NS mass distribution, while difficult to measure at present due to small numbers, contains a wealth of information about the outcomes of stellar evolution, the CCSN mechanism, and the nuclear EOS \citep{raithel:2018}. 
In Figure \ref{f.nsdist} we show the IMF-weighted NS mass distribution from our STIR simulations for several values of \alphL. 
The bars in this plot are color-coded based on the ZAMS mass of the progenitor from whence the NSs at that mass came.

In \citet{raithel:2018} the authors compute the NS mass distribution from the 1D explosion models of \citet{sukhbold:2016}.
They find that a roughly Gaussian distribution, peaked around a gravitational mass of 1.4 \Msun fits the models quite well. 
This is also in rough agreement with the observed average NS mass (Figure \ref{f.avgnsmass}).
There is tentative evidence, however, based on the population of millisecond pulsars that the NS mass distribution is double-peaked, with a primary peak around a gravitational mass of 1.4 \Msun and a secondary peak around a gravitational mass of 1.8 \Msun \citep{antoniadis:2016}. 
The models of \citet{sukhbold:2016} have trouble explaining such a high mass peak, if it exists \citep{raithel:2018}. 
In Figure \ref{f.nsdist} we also plot the double-Gaussian model of \citet{antoniadis:2016} based on masses derived from millisecond pulsars. 

At $\alphL=1.2$ (top left panel of Figure \ref{f.nsdist}), the PNS mass distribution is peaked at small gravitational mass, around 1.25 \Msun, with a second peak around 1.8 \Msun.
This structure is a result of only the least compact and most compact progenitors exploding at this low \alphL value. 
As \alphL is increased, the NS mass distribution fills in with more and more progenitors successfully exploding.
At $\alphL=1.25$, the distribution peaks at a gravitational mass of about 1.4 \Msun with a clear second peak at about 1.8 \Msun, remarkably similar to the double Gaussian model of \citet{antoniadis:2016}. 
As \alphL increases beyond 1.25, more models explode and explosions occur earlier for already-exploding progenitors. 
As stars in the mass range 13-16 \Msun begin exploding, the peak in the NS mass distribution shifts to slightly larger values. 
By $\alphL=1.4$, the double-peaked structure of the distribution remains, but the first Gaussian peak is far more dominant and peaks around 1.7 \Msun gravitational, a bit larger than the peak estimated by \citet{antoniadis:2016} and \citet{raithel:2018}. 
For no value of \alphL do we find very massive neutron stars, near 2 \Msun gravitational. 
This could be a product of explosions occurring too early in STIR, before the PNS has grown in mass sufficiently, or due to a deficiency in the set of progenitor models we use. 

We assume all models that fail to explode result in gravitational collapse of the PNS to a BH. 
For progenitors with compactnesses $\xi_{2.5} \lesssim 0.2$, our simulations run late enough to capture the onset of PNS collapse directly.
For all other models that do not explode within about 5\,s, we consider them to also be failed explosions resulting in BH formation.
For estimating the mass of the BH formed by a given progenitor, we use the approach presented by \citet{fernandez:2018}.
Here the authors account for unbinding of some fraction of the envelope of the star due to neutrino mass loss from the core prior to BH formation \citep[e.g.,][]{lovegrove:2013}.
While a rather detailed model, it yields estimates for BH masses fairly close to the He core mass of the star at the point of initial core collapse \citep[cf.][]{clausen:2015}.

In Figure \ref{f.bhdist} we show the IMF-weighted distributions of BH masses derived from our STIR simulations for a few values of \alphL. 
As for the NS mass distributions (Figure \ref{f.nsdist}), the BH mass distributions are color-coded according to the ZAMS mass of the BH's progenitor.
At \alphL = 1.0 (left panel), all progenitors with ZAMS masses $\gtrsim$10 \Msun fail to explode and will form BHs.
At $\alphL=1.2$, some of the most compact progenitors in the range of 20 to 30 \Msun explode, along with more of the lowest-mass models. 
This removes BHs in the 5.0 to $\sim$9.0 \Msun range.
About 73\% of massive stars explode for $\alphL=1.25$, leaving a more anemic distribution of BHs. 
The right panel of Figure \ref{f.bhdist} clearly shows that STIR predicts a typical BH mass of around 4 \Msun, with the vast majority of those BHs originating from stars with ZAMS masses in the 13 to 16 \Msun range. 
The distribution is fairly tight from 3 \Msun to 7 \Msun with a low-probability tail extending up to 10 \Msun resulting from higher ZAMS mass progenitors.
It is interesting to note that for $\alphL = 1.25$, the BHs we find come from progenitors with ZAMS masses representing almost the entire range of high mass stars, from 12 \Msun to 60 \Msun. 
Clearly, according to STIR, forming black holes is not something only the highest mass stars do. 
As \alphL increases, more progenitors explode and the distribution of BH mass tightens around the same mean value of 5 \Msun. 

For this progenitor set, we find no BHs with masses above about 15 \Msun for any value of \alphL following the prescription for predicting the BH mass of \citet{fernandez:2018}. 
Thus our present model set based on single, solar-metallicity stars would have difficulty explaining the extremely massive BHs now being observed in gravitational waves by LIGO \citep[e.g.,][]{abbott:2016}.
Due to the mass loss prescription used in this progenitor set, even if we were to assume that the entire star collapsed into to BH, without any unbinding of the outer envelope as envisaged by \citet{lovegrove:2013}, this would still be insufficient to produce the 30-\Msun and more BHs observed so far by LIGO \citep[cf.][]{ebinger:2019}.
As has been pointed out by other works \citep{belczynski:2016}, forming such enormous BHs via single star evolution is a challenge, if not an impossibility. 

Taken together, our NS and BH mass distributions show only weak evidence for the presence of a ``mass gap'' between NSs and BHs. 
\citet{belczynski:2012} and \citet{wiktorowicz:2014} present observational and theoretical support for the existence of such a gap between about 2 \Msun for the highest mass NSs and 5 \Msun for the lowest mass BHs, i.e., no BHs with baryonic masses less than about 5 \Msun. 
The data seem to show that for BHs that have measured masses so far, many of them ``pile up'' around 5 \Msun, with no statistically significant evidence that any BHs fall below this mass \citep{wiktorowicz:2014}. 
For our model set we find a substantial number of BHs below 5 \Msun for all values of \alphL. 
This is a product of STIR predicting failed explosions for certain lower-mass progenitors in the range of 10 \Msun to 19 \Msun. 
These stars yield fairly low mass BHs according to the method we have adopted for estimating final BH masses \citep{fernandez:2018}.

\subsection{Comparison to 2D and 3D simulations}
\label{s.comparison}

As discussed above, our results with STIR reproduce the qualitative result from other parameterized 1D models \citep[e.g.,][]{ugliano:2012, ertl:2016,sukhbold:2016, ebinger:2019} that the explodability of massive stars is a complicated, non-monotonic function of ZAMS mass. 
We also find, however, that the details of which precise stars explode or fail is quite different from other 1D parameterizations, specifically that of \citet{ugliano:2012}; \citet{sukhbold:2016}; and \citet{ebinger:2019}.
So, which is ``right?''
Well, this is a complicated question to answer and depends on how one approaches addressing it.
In this section, we compare the explosions and failures found by STIR to some of the available results in the literature from multidimensional CCSN simulations including comparable physics. 
We also compare these results to those two purely neutrino-driven 1D explosion parameterizations: \citet{sukhbold:2016} and \citet{ebinger:2019}. 

One approach to determining the verisimilitude of a 1D explosion model would be to compare the population statistics of the produced set of models to the population of observed CCSNe.
1D parameterizations such as those of \citet{sukhbold:2016} and \citet{ebinger:2019} actually do quite well in reproducing mean observable metrics such as the explosion fraction, explosion energy, radioactive nickel production, and even detailed nucleosynthesis \citep{curtis:2019}.
These methods, however, are tuned to accurately yield these metrics for certain progenitors (usually those designed to mimic SN 19987A). 
For STIR, we make no such fitting to observable outcomes but instead ``fit''  the main model parameter, \alphL, to the 3D simulation of \citet{oconnor:2018b} while fixing all the diffusion parameters to fiducial values (see Section \ref{s.stir3d}). 
This comparison showed that a mixing length parameter of \alphL = 1.2 - 1.3 worked well to reproduce the strength and location of turbulent convection and shock radius evolution of the 3D model with STIR. 
Consideration of the predicted explosion fraction for the population of CCSNe produced from the \citet{sukhbold:2016} progenitors using STIR (Section \ref{s.expl}), led us to prefer an \alphL value of 1.25, at least amongst those values we simulate in this paper.
We will stick with this value of \alphL for our comparisons in this subsection, except where noted.

We first consider how our results with STIR compare to the 2D simulations of \citet{oconnor:2018}. 
The neutrino transport used in that work is exactly the same code we use in this work (two-moment M1), modulo the diffusion of trapped neutrino fractions due to turbulent convection (Section \ref{s.num}). 
In \citet{oconnor:2018} the authors simulated progenitors of ZAMS masses 12, 15, 20, 21, 22, 23, 24, and 25 \Msun, all of which were taken from \citet{woosley:2007}. 
Here we employ the models from \citet{sukhbold:2016} but comparison shows these two sets to be fairly similar for the masses we are concerned with. 
\citet{oconnor:2018} also use the \citet{lattimer:1991} with nuclear incompressibility of 220 MeV (hereafter LS220), where here we use the SFHo EOS \citep{steiner:2013}. 
This could be a source of quantitative difference in the results but the controlled 2D comparison between these two EOSs by \citet{pan:2019} implies that they are qualitatively very similar and, crucially, should agree on the leading-order question of whether or not a given progenitor explodes. 

\begin{figure}[tb]
  \centering
  \includegraphics[width=0.45\textwidth]{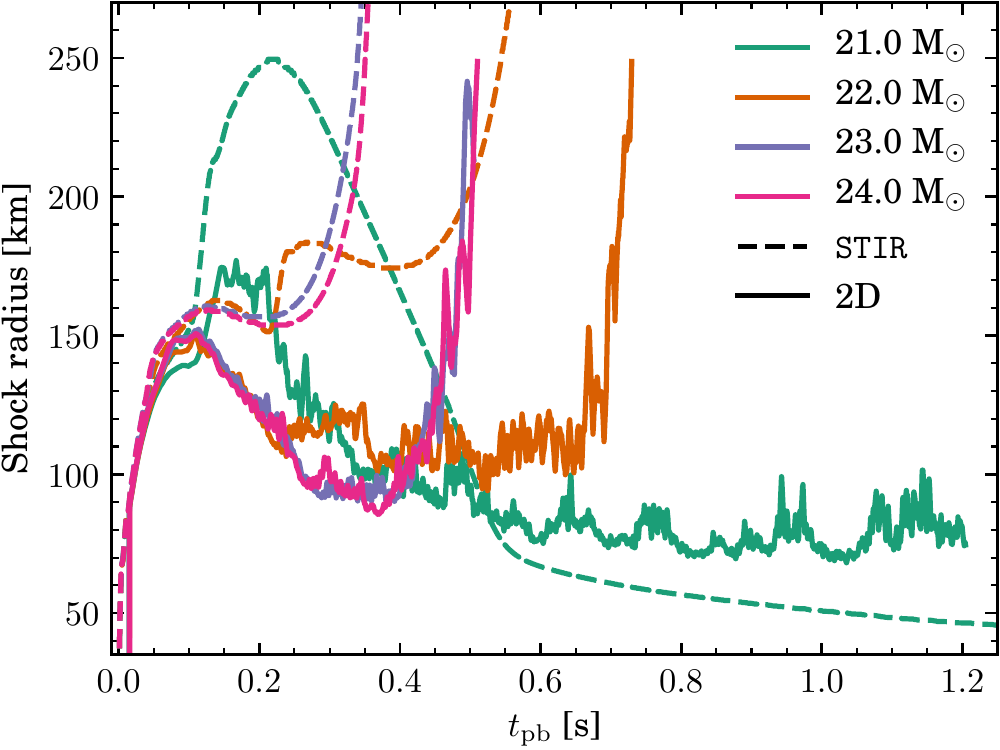}
  \caption{Shock radius curves for select STIR simulations in the 21.0-, 22.0-, 23.0-, and 24.0-\Msun progenitors for \alphL = 1.2 (dashed lines). Also shown are the shock radius curves for the same ZAMS mass progenitors from the 2D simulations of \citet{oconnor:2018} (solid lines).}
  \label{f.stirVs2D}  
\end{figure}

\begin{deluxetable*}{lr|ccccccccccccc}[!t]
    \tablecaption{Explosion times from recent studies in 1D, 2D, and 3D using similar methods and progenitors compared to STIR with \alphL = 1.25. \label{t.explosions}}
    \tablehead{
    \colhead{Progenitor Mass [\Msun]:} & \colhead{} & \colhead{9.0} & \colhead{10.0} & \colhead{11.0} & \colhead{12.0} & \colhead{13.0} & \colhead{14.0} & \colhead{15.0} & \colhead{16.0} & \colhead{17.0} & \colhead{18.0} & \colhead{19.0} & \colhead{20.0 } & \colhead{25.0} 
    }
    \startdata
    STIR (this work)                       & 1D  & 0.24 s & 0.40 s & 0.29 s & 0.40 s  & \nodata & \nodata & \nodata & 0.43 s & 0.39 s & 0.64 s & 0.47 s & 0.40 s & 0.39 s  \\
    \citet{sukhbold:2016}\tablenotemark{a} & 1D  & 0.66 s & 0.80 s & 1.56 s &  1.1 s  & 0.8 s   & 0.71 s  & \nodata & 0.78 s & 0.77 s & 1.2 s  & 1.2 s  & \nodata& \nodata \\
    \citet{ebinger:2019}\tablenotemark{b}  & 1D  &        &        & 0.39 s & 0.47 s  & 0.41 s  & 0.41 s  & 0.44 s  & 0.35 s & 0.33 s & 0.42 s & 0.36 s & 0.42 s & 0.49 s  \\
    \citet{oconnor:2018}\tablenotemark{c}  & 2D  &        &        &        & \nodata &         &         & 1.1 s   &        &        &        &        & 0.8 s  & 0.6 s   \\
    \citet{vartanyan:2018}\tablenotemark{d}& 2D  &        &        &        & \nodata & \nodata &         & \nodata & 0.3 s  & 0.38 s &        & 0.41 s & \nodata& \nodata \\
    \citet{summa:2016}\tablenotemark{e}    & 2D  &        &        &        & 0.79 s  &         &         & 0.62 s  &        &        &        &        & 0.35 s& 0.40 s  \\
    \citet{bruenn:2016}\tablenotemark{f}   & 2D  &        &        &        & 0.24 s  &         &         & 0.23 s  &        &        &        &        & 0.21 s & 0.21 s  \\
    \citet{pan:2018}\tablenotemark{g}      & 2D  &        &        &        &         &         &         &         &        &        &        &        &        &         \\
    \citet{ott:2018}\tablenotemark{h}      & 3D  &        &        &        & \nodata &         &         & 0.58 s  &        &        &        &        & 0.35 s &         \\
    \citet{lentz:2015}\tablenotemark{i}    & 3D  &        &        &        &         &         &         & 0.37 s  &        &        &        &        &        &         \\
    \citet{melson:2015a}\tablenotemark{j}  & 3D  &        &        &        &         &         &         &         &        &        &        &        & \nodata&         \\
    \citet{burrows:2019a}\tablenotemark{d} & 3D  & 0.21 s & 0.45 s & 0.21 s & 0.31 s  & \nodata & \nodata & \nodata & 0.30 s & 0.30 s & 0.32 s & 0.40 s & 0.45 s & 0.55 s         
    \enddata
    \tablenotetext{a}{Same progenitor models as this work; parameterized neutrino light bulb; no nuclear EOS.}
    \tablenotetext{b}{\citet{woosley:2007} progenitors; parameterized ``PUSH'' IDSA transport; DD2 EOS.}
    \tablenotetext{c}{\citet{woosley:2007} progenitors; same transport as this work; LS220 EOS.}
    \tablenotetext{d}{\citet{woosley:2007} progenitors; M1 transport with slightly different neutrino microphysics; SFHo EOS.}
    \tablenotetext{e}{\citet{woosley:2007} progenitors; two-moment VET transport; LS220 EOS.}
    \tablenotetext{f}{\citet{woosley:2007} progenitors; FLD transport; LS220 EOS.}
    \tablenotetext{g}{\citet{woosley:2007} progenitor; IDSA transport; LS220 EOS.}
    \tablenotetext{h}{\citet{woosley:2007} progenitors; M1 transport without velocity-dependent terms; SFHo EOS; fully general realistic.}
    \tablenotetext{i}{\cite{woosley:2007} progenitor; FLD transport; LS220 EOS.}
    \tablenotetext{j}{\citet{woosley:2007} progenitor; two moment VET transport; LS220 EOS; note that this model explodes around 0.4 s if the authors include matter strangeness corrections.}
    \tablecomments{Explosion times are roughly approximate based on published data. A ``\nodata'' mark implies that the model did not explode in the simulated time and may, or did, collapse to a BH. A blank entry implies that a given progenitor model was not simulated by the referenced work. All referenced models (except \citet{sukhbold:2016}) use some form of general relativistic gravity. }
  \end{deluxetable*}

In Figure \ref{f.stirVs2D} we show the shock radius curves for 21-, 22-, 23-, and 24-\Msun progenitors for the 2D simulations of \citet{oconnor:2018} along with the comparable 1D STIR simulations (with \alphL = 1.2).
This is an interesting set of progenitor masses since \citet{oconnor:2018} find that the 21-\Msun model fails while the others succeed with the 22-\Msun exploding 200 ms later than the 23- and 24-\Msun models. 
Encouragingly, STIR predicts the exact same qualitative outcome: the 21-\Msun progenitor fails while the others succeed in exploding. 
The order of explosion times is also the same with the 22-\Msun star exploding the latest, but all of the explosion times are significantly earlier as compared with the 2D simulations.
This could be a result of the differences in EOS or overly optimistic mixing parameters in STIR.
We note that for $\alphL=1.23$ and above, STIR predicts that all of these progenitors explode. 

In Table \ref{t.explosions} we compare the explosion times for our STIR simulations to several recent 1D, 2D, and 3D simulations of CCSNe using similar physics and progenitors. 
For brevity, we show only the progenitors models simulated in 3D as summarized in \citet{burrows:2019a}.
In this Table, failed explosions (or, more precisely, a lack of explosion within the simulated time) is represented by ``\nodata.''
As discussed above for the cases of 21- to 25-\Msun stars, STIR also agrees fairly well with the 2D results of \citet{oconnor:2018}.
Table \ref{t.explosions} shows the results for the other models simulated by \citet{oconnor:2018}.
For the 20.0- and 25.0-\Msun progenitors, STIR agrees well with the 2D results, though the explosion times are earlier in STIR.
For the 12- and 15-\Msun models, STIR and \citet{oconnor:2018} disagree on the qualitative outcome.
The 12-\Msun model explodes in STIR but fails in \citet{oconnor:2018} and vice versa for the 15-\Msun progenitor.
We also show in Table \ref{t.explosions} the approximate explosion times for these progenitors from the 1D parameterized purely neutrino-driven simulations of \citet{sukhbold:2016}. 
In every instance, \citet{sukhbold:2016} and \citet{oconnor:2018} disagree on the qualitative outcome of explosion or not for these progenitors.
Table \ref{t.explosions} also shows the explosions times from \citet{ebinger:2019} for these progenitor masses from the 1D PUSH model. 
Here, the authors use the \citet{woosley:2007} progenitors and the DD2 EOS parameterization \citet{fischer:2014}. 
\citet{pan:2019} compare DD2 to SFHo in 1D and find that it produces systematically larger shock radii and, therefore, might be more favorable for explosion, but for the most part gives similar results to SFHo. 
PUSH is, apparently, very robust and almost all of these progenitors successfully explode and at times earlier than either \citet{sukhbold:2016} or STIR. 
For the progenitors used by \citet{oconnor:2018}, the only models that PUSH predicts failure for are the 23-\Msun and 24-\Msun stars. 
These stars are the first to explode in \citet{oconnor:2018}, and also for STIR (see Figure \ref{f.stirVs2D}).

The 2D explosions of \citet{summa:2016} also show fairly good agreement with STIR. 
There the authors simulate the 12-, 15-, 20-, and 25-\Msun progenitors from \citet{woosley:2007} using a two-moment variable Eddington tensor (VET) method with model Boltzmann equation closure for neutrino transport with detailed microphysics  \citep{rampp:2002} and the LS220 EOS.
They find explosions for all of these progenitors, in agreement with \citet{oconnor:2018} except for the case of the 12-\Msun progenitor. 
STIR also predicts explosion for the 12-\Msun progenitor.
For this case, however, \citet{summa:2016} find the latest explosion of their entire set. 
\citet{bruenn:2016} also find explosions in their 2D simulations for all the same progenitors as \citet{summa:2016}. 
There, the authors employ one-moment flux-limited diffusion (FLD) neutrino transport, the LS220 EOS, and include a transition to a full nuclear reaction network at low-density \citep{bruenn:2018}.
\citet{bruenn:2016} find that all of these progenitors explode around the same time, just after 200 ms post-bounce. 
This qualitatively agrees with STIR, though the spread of explosion times from STIR is a bit larger, and quite comparable to the explosion times for the 3D simulation summarized in \citet{burrows:2019a}.

\citet{vartanyan:2018} present 2D results for several of the \citet{woosley:2007} progenitors, including all of those used by \citet{bruenn:2016} and \citet{summa:2016}. 
The explosion times they find are summarized also in Table \ref{t.explosions}. 
\citet{vartanyan:2018} employ a two-moment M1 neutrino transport approach similar to ours \citep{skinner:2018} but with significantly different neutrino-matter interactions \citep{burrows:2018}.
They also use the SFHo EOS. 
Their results for the progenitors considered are similar to what STIR predicts, with a few exceptions.
STIR and \citet{vartanyan:2018} disagree on the qualitative outcome of explosion or failure for the 12.0-, 20.0-, and 25.0-\Msun progenitors. 
All of the other 2D simulations they present yield the same conclusion on this most basic question as compared with STIR. 
We note, however, this puts the results of \citet{vartanyan:2018} in tension with the other recent 2D simulations we mention above \citep{summa:2016, bruenn:2016, oconnor:2018} and, in some cases, even with the 3D results using the same code \citep{vartanyan:2019, radice:2018c, burrows:2019a}.

Given the issues that affect the physical accuracy of 2D simulations, most relevantly the incorrect inverse turbulent energy cascade \citep{hanke:2012, couch:2013a,murphy:2013, couch:2014, couch:2015a}, comparing STIR to fully 3D simulations is desirable. 
There are still only a handful of 3D CCSN simulations that include high-fidelity neutrino transport and microphysics in the literature.
A large set of comparable 3D simulations has been completed using the FORNAX code and discussed in \citet{vartanyan:2019, radice:2018a, burrows:2019a}. 
In every single case, STIR and \citet{burrows:2019a} agree on the qualitative outcome of explosion or failure.
This is true even for the progenitors in the mass range 13 to 15 \Msun which fail both for STIR and in \citet{burrows:2019a}.
The explosion times found in these 3D simulations are also relatively similar to those found in STIR.
Considering the purely neutrino-driven parameterization of \citet{sukhbold:2016}, their 1D models disagree with the 3D simulations presented in \citet{burrows:2019a}, and hence with STIR, for the cases of the 13.0-, 14.0-, 20.0-, and 25.0-\Msun models. 

In \citet{ott:2018} the authors present a set of 3D general relativistic CCSN simulations using M1 neutrino transport with the 12-, 15-, 20-, and 40-\Msun progenitors of \citet{woosley:2007}. 
For these models, \citet{ott:2018} find explosions for the 15-, 20-, and 40-\Msun stars, and a failure for the 12-\Msun progenitor. 
Their explosion times are also listed in Table \ref{t.explosions}.
As mentioned above, STIR predicts failure for the 15-\Msun star but explosion for the 12-\Msun, in contrast to the 3D results of \citet{ott:2018}. 
STIR also finds a successful explosion for the highly-compact 40-\Msun model.
As mentioned above, depending on the EOS employed, \citet{pan:2019} also find successful explosions for this progenitor, though accompanied by simultaneous BH formation.
It is possible that STIR would predict a ``fallback'' BH formation after this successful explosion if we continued our simulations to much later times  \citep{chan:2018}. 
Table \ref{t.explosions} shows a few other 3D simulations with comparable physics to STIR. 
\citet{lentz:2015} present the successful explosion in 3D of the 15-\Msun progenitor from \citet{woosley:2007} using the same simulation method as in \citet{bruenn:2013, bruenn:2016}. 
The explosion for this progenitor occurs later than for \citet{bruenn:2016}, but nonetheless still occurs, in disagreement with the prediction from STIR of failure for this model. 
\citet{melson:2015a}, using a similar approach as that of \citet{summa:2016}, find a failed explosion for the 20-\Msun progenitor of \citet{woosley:2007} unless strange-quark corrections are accounted for in the neutrino-nucleon scattering cross sections. 
STIR predicts an explosion for this model.

\section{Conclusions} \label{s.conc}

We have presented a new approach for incorporating the effects of convection and turbulence into 1D simulations of the CCSN mechanism called Supernova Turbulence In Reduced-dimensionality (STIR). 
STIR begins with a Reynolds decomposition of the compressible Euler equations into background, mean flow components and perturbed, turbulent components (see Section \ref{s.turb}). 
We then angle-average these equations in spherical coordinates, and make a few simplifying assumptions gleaned from multidimensional simulations of CCSNe, to yield a set of 1D evolution equations that depend solely on the local magnitude of the turbulent kinetic energy. 
We ``close'' these equations using a modified MLT approach that relates the space- and time-dependent evolution of the turbulent kinetic energy to the local \brunt frequency (Equation (\ref{e.fbv2})). 
This physically-motivated model then depends on a total of five free parameters, the mixing length parameter \alphL (Equation (\ref{e.Lamm})) and four parameters that control the strength of diffusive mixing due to turbulent convection for internal energy ($\alpha_e$), turbulent kinetic energy ($\alpha_K$), electron fraction ($\alpha_{Y_e}$), and trapped neutrino fractions ($\alpha_\nu$).
In the present work, we vary only the mixing length parameter \alphL and fix all the diffusive mixing parameters to 1/3 \citep{bruenn:1995}. 

In STIR, we use full high-fidelity two-moment M1 neutrino transport \citep{oconnor:2015, oconnor:2018} and make no modifications to the transport or neutrino microphysics, beyond the inclusion of diffusive mixing of trapped neutrino fractions due to convection.
We also include the full PNS on the computational domain and utilize a microphysical nuclear EOS. 
These features of STIR allow us to explore the dependence of the CCSN mechanism and resulting observables to details of the nuclear and neutrino physics. 
In future work, we will explore these aspects further.

In Section \ref{s.stir3d} we compare STIR to the 3D simulation from \citet{oconnor:2018b} using the very same neutrino transport code we employ in STIR. 
By varying only \alphL, STIR can reasonably reproduce the angle-averaged features and dynamics of this 3D model, including the profiles of turbulent motions, electron fraction, and entropy as well as the time evolution of the average shock radius. 
The 1D STIR model compares far better to the full 3D simulation than does a 1D simulation that neglects turbulent convection entirely (i.e., \alphL = 0.0). 
The value of \alphL we find that fits best to the dynamics of this particular 3D simulation is between 1.2 and 1.3.
As reported by \citet{oconnor:2018b}, this 3D simulation fails to explode and STIR also predicts failure for this model up to \alphL = 1.25.
For sufficiently large values of \alphL, turbulence-aided neutrino-driven explosions are achieved.

Comparison to the 3D simulation of \citet{oconnor:2018b} showed that STIR does a comparatively poor job of capturing convection in the PNS. 
The ability of STIR to model PNS convection could be improved by including the effects of trapped neutrino fraction gradients in our calculation of the \brunt frequency. 
Equation (\ref{e.fbv2}) implicitly includes only the electron fraction gradient and not the full lepton fraction gradient that would include the trapped neutrino fractions, as well. 
Calculating this correctly is complicated, but may be critical to correctly modeling PNS convection with STIR \citep{roberts:2012a}. 

In Section \ref{s.param}, we present a first, preliminary parameter study of observable outcomes for a population of CCSNe generated by STIR in progenitors of masses 9 to 120 \Msun from \citet{sukhbold:2016}. 
We find a similar pattern of ``islands of explosion'' described by \citet{sukhbold:2016} wherein the explodability of the progenitors is a complicated, non-monotonic function of their ZAMS masses. 
With STIR, however, the details of which precise progenitor models explode and which fail is quite different from other 1D explosion parameterizations, such as \citet{sukhbold:2016} and \citet{ebinger:2019}. 
We compare our results for STIR for several specific progenitor masses to recent results from 2D and 3D simulations in Section \ref{s.comparison}. 
We find tentative evidence based on certain sets of 2D and 3D simulations that STIR predicts more accurately which progenitors will explode or not than other 1D models \citep{sukhbold:2016,ebinger:2019}, but much more work is needed both with STIR and multidimensional simulations to say anything definitive along these lines. 

The total fraction of massive stars that explodes according to STIR is, unsurprisingly, a strong function of \alphL (see Section \ref{s.expl}). 
For \alphL = 1.25, we find an IMF-weighted explosion fraction of massive stars above 9 \Msun of about 73\%, in decent agreement with the explosion fraction estimated by \citet{adams:2017}. 
We also find a population-averaged diagnostic explosion energy of around 0.7$\times$10$^{51}$ erg. 
The explosion energies in most of our models are still increasing at the end of our simulations (see Figure \ref{f.alpha_ener}) and so the explosion energies we report here are lower limits. 
In turn, though, we neglect accounting for the ``overburden'' energy to unbind the outer layers of the progenitors \citep{bruenn:2016}. 
The overburden would revise the final explosion energies downward. 

In Section \ref{s.criteria} we compare the results from STIR to a few explosion criteria for predicting success or failure based on features of the precollapse progenitor models. 
We find essentially no correlation between the ``critical'' \alphL value required to just achieve an explosion and the progenitors' compactness $\xi_{1.75}$ (Equation (\ref{e.compact})).
The highest compactness models ($\xi_{1.75}\gtrsim0.8$) explode more readily than lower compactness progenitors, though so do the very lowest compactness cases.
There is some weak evidence from multidimensional simulations that higher compactness progenitors also explode more readily (see Section \ref{s.comparison}). 

We also compare our results to the two-parameter explosion criterion of \citet{ertl:2016}. 
We find a very different separation between explosions and failures in the $\mu_4$-$M_4 \mu_4$ plane (Equations (\ref{e.mu4}) and (\ref{e.m4})) than do \citet{ertl:2016}. 
There does not seem to be an easy way of reconciling this difference with, e.g., different parameters for their separation curve. 
This reflects the fundamental differences in the natures in which explosions are driven between turbulence-aided STIR and the purely-neutrino driven methods such as \citet{ertl:2016} and \citet{sukhbold:2016}. 

A key observable outcome from CCSN explosions is the mass distributions of the compact remnants, NSs and BHs, they leave behind. 
In Section \ref{s.remnants} we construct predicted NS and BH mass distributions based on STIR and compare them to what available observational data there are for these quantities. 
The average NS mass is sensitive to the \alphL parameter, at first because a very different fraction of the progenitors we simulate explode as \alphL is increased, and then because increasing \alphL causes explosions to occur earlier and earlier. 
In our 1D model, once an explosion sets in, accretion onto the PNS is stopped, in contrast to multidimensional models that can exhibit simultaneous explosion and continued accretion. 
Thus, the later the explosion with STIR, the heavier the resulting PNS. 
In general, at our preferred \alphL of 1.25, STIR predicts a PNS mass distribution and the IMF-weighted average NS mass in good agreement with the current observational estimates.

This first parameter study with STIR is, in large part, a proof of principle. 
There are several aspects of our model that can be improved, and we will do so in future work. 
First, we must relax the assumption of NSE everywhere.
This will require the transition to an appropriate low-density EOS and some sort of nuclear network to track and evolve the isotopic composition of the plasma. This alone would be a major improvement and allow us to push to much later simulation times, more accurately simulating, e.g., the final diagnostic explosion energies.
By assuming NSE in the low density material that is, in actuality, far from NSE we lose the energy that could be derived from the nuclear processing of this material toward NSE. 

In future work we will also explore a more accurate handling of the full lepton gradient in computation of the \brunt frequency. 
This could improve the behavior of PNS convection in our STIR simulations.
We also plan a statistically more rigorous fitting of the five STIR ``$\alpha$'' parameters to available 3D data and will explore the universality of these parameters for different progenitor stars.

Another interesting potential for STIR is to include turbulent convection in the precollapse progenitor star. 
Our evolution equation for the turbulent kinetic energy, Equation (\ref{e.turbk}), implies that in the layers above the shock where the \brunt frequency should be close to zero, turbulent kinetic energy will be advected along with the collapsing stellar core. 
So, the convective structure of the progenitor could be fully retained in our STIR simulations, likely resulting earlier and easier explosions \citep{couch:2013b, muller:2015, muller:2016a, couch:2015}.
In STIR, this would manifest itself as providing a large, finite amplitude initial perturbation from which post-shock neutrino-driven convection would grow, ultimately increasing the overall strength of post-shock turbulence and convection (see Equation (\ref{e.vtrbmin}) and surrounding discussion). 
In addition, the evolution equation for the turbulent kinetic energy in spherical coordinates implies a geometric concentration of the turbulent kinetic energy as precollapse perturbations are advected to smaller radii.
This results in an amplification of such pre-existing perturbations during collapse which is essentially equivalent to that predicted by \citet{lai:2000}.  

It is worth pointing out that while STIR seems a promising model for theoretical studies of CCSN populations and their resulting observables it is  hampered by the fact that it is fundamentally a 1D model for a quintessentially 3D phenomenon.  
Crucially, while the importance of the SASI for CCSN {\it explosions} is still much-debated \citep{hanke:2012, murphy:2013, fernandez:2014, cardall:2015, summa:2018, vartanyan:2019}, it is a real instability and can occur in the 3D CCSN context \citep[e.g.,][]{oconnor:2018b}. 
The impact of the SASI, whatever it may be, is something that is not accounted for in STIR but is something that we should expect in 3D simulations.
In other words, real CCSNe are both {\it shaken} by the SASI and {\it stirred} by turbulent convection. 

Current missions in time-domain astronomy, such as the All-Sky Automated Survey for Supernovae and the Zwicky Transient Facility, are already revolutionizing the observational picture of CCSNe.
Future missions such as the Large Synoptic Survey Telescope will completely redefine what the astronomical study of CCSNe even means through a cataclysm of new data.
These efforts will fill out our understanding of the real CCSN {\it population}. 
In order to translate these data into understanding, we need a comparably strong theoretical picture. 
This necessitates high-fidelity simulations of CCSNe from the entire range of possible progenitors and the production of reliable predictions for CCSN observables that can be compared directly with data.
High-fidelity 3D simulations of CCNSe are still incredibly difficult and expensive, limiting our ability to cover the enormous parameter space of initial conditions.
In the meantime, 1D parameterized explosion models offer an attractive alternative because their relative affordability means that thousands of simulations can be run quickly. 
Still, it is crucial that such 1D models accurately reproduce key features of 3D simulations. 
By including a physically rigorous model for the impact of convection and turbulence in 1D, STIR is a promising approach for making testable theoretical predictions about the broad and complex CCSN population.

\acknowledgments
The authors would like to thank Ed Brown, Alex Heger, Grant Mathews, Jeremiah Murphy, and Tuguldur Sukhbold for useful discussions that influenced this work.
We thank Luke Roberts for useful discussions regarding PNS convection and for providing data on the gravitational mass for the SFHo EOS.
SMC is supported by the U.S. Department of Energy, Office of Science, Office of Nuclear Physics, under Award Numbers DE-SC0015904 and DE-SC0017955 and the Chandra X-ray Observatory under grant TM7-18005X.
MLW is supported by the U.S. Department of Energy, Office of Science, Office of Nuclear Physics, under Award Number DE-SC0015904 and by an NSF Astronomy and Astrophysics Postdoctoral Fellowship under award AST-1801844.
EO is supported by the Swedish Research Council under Award Number 2018-04575.
This work was supported in part by Michigan State University through computational resources provided by the Institute for Cyber-Enabled Research.

\software{
  \href{http://flash.uchicago.edu/site/}{FLASH} \citep{fryxell:2000,dubey:2009}, 
  \href{https://matplotlib.org/}{Matplotlib} \citep{hunter:2007a}, 
  \href{http://www.numpy.org/}{NumPy} \citep{walt:2011}, 
  \href{https://www.scipy.org/}{SciPy} \citep{jones:2001},
  \href{http://www.nulib.org/}{NuLib} \citep{oconnor:2015}
}
\\
\vspace{0.1in}

\bibliography{STIR}

\end{document}